\documentclass[11pt(*)]{JHEP3}
\usepackage{appendix}
\usepackage{amsmath}
\usepackage{graphicx}
\usepackage{amssymb}
\usepackage{amsthm}
\usepackage{latexsym}
\usepackage{mathrsfs}
\usepackage{slashbox}
\usepackage[square, comma, sort&compress, numbers]{natbib}

\newcommand\fverb{\setbox\fverbbox=\hbox\bgroup\verb}
\newcommand\fverbdo{\egroup\medskip\noindent%
	    \fbox{\unhbox\fverbbox}\ }
\newcommand\fverbit{\egroup\item[\fbox{\unhbox\fverbbox}]}
\newbox\fverbbox

\title{Particle with non-Abelian charge: classical and quantum}
\author{Amitabha Lahiri, Tae-Hun Lee\\
\email{amitabha@bose.res.in}, \email{taehunl@bose.res.in}\\
Department of Theoretical Sciences\\
S.N. Bose National Centre for Basic Sciences\\
Block JD, Sector III, Salt Lake, Kolkata, 700 098, W.B., India.}

\abstract{We study the action for a non-Abelian charged particle in
  a non-Abelian background field in the worldline formalism,
  described by real bosonic variables, leading to the well known
  equations given by Wong. The isospin parts in the action can be
  viewed as the Lagrange multiplier term corresponding to a
  non-holonomic constraint restricting the isospins to be parallel
  transported. The path integration is performed over the isospin
  variables and as a result, the worldlines turn out to be
  constrained by the classical solutions for the isospins.

  We derive a wave equation from the path integral, constructed as
  the constrained Hamiltonian operator acting on the wave
  function. The operator ordering corresponding to the quantum
  Hamiltonian is found and verified by the inverse Weyl transformation.}

\begin{document}
\section{Introduction}
A starting point in the usual way of quantization for a system is
to adopt its classical system as a fundamental
framework. Interestingly, different from the case of an
electrically charged particle, the dynamics of a particle with
non-Abelian charge was quantized directly from analogy and
extension of its quantum theory since a non-Abelian charged
particle had not been observed in a classical context. If a
classical dynamics of a particle is considered as the most
fundamental basis to construct a quantum theory, to find a
classical picture of a particle with non-Abelian charge can be not
only an interesting quest but also a crucial part for understanding
its quantum nature.

A set of classical equations of motion for a non-Abelian charged
particle was first derived by Wong from the quantum action in which
a non-Abelian charged particle is described by a Dirac field
\cite{Wong:1970fu}. These are called Wong's equations and include
the parallel transport equation describing an isospin in addition
to a non-Abelian extension of the Lorentz force equation.

The purpose of this paper is to find a classical picture for a
non-Abelian charged particle in a background field. We start from a
classical action for a non-Abelian charged particle and construct
the path integral. Many actions producing the Wong equations have
been proposed \cite{Balachandran:1976ya, Balachandran:1977ub,
  Duval:1981js, Horvathy:1982cs}. Since we are interested in a
purely classical picture, we write the action using only variables
valued in the real numbers.  In particular, all our $c$-numbers
commute with one another, so upon quantization they will become
bosonic variables.  In \cite{Balachandran:1976ya}, using the
constrained Hamiltonian formalism, a wave equation governed by the
constrained Hamiltonian operator was written down though the
operator orderings were unspecified.
We find that the wave equation gets an additional term 
which breaks gauge invariance. This will be verified by an 
inverse Weyl transformation. The derivation brings forth 
certain issues of operator ordering.

As is usually done, for instance in \cite{Balachandran:1976ya,
  Balachandran:1977ub}, we first consider the equations given by
Wong as a starting point to eventually obtain a quantum description
for a non-Abelian charged particle. The first step is to obtain the
classical action producing these equations.  As mentioned above, we
use only commuting real variables for all degrees of freedom.  Next
we quantize this action using the constrained Hamiltonian formalism
due to Dirac \cite{Dirac:1967, Henneaux:1992ig}. Once the
Hamiltonian is obtained, we construct the path integral in the
worldline formalism, and derive the constrained wave equation. This
equation is compared with the one derived in canonical quantization
earlier, discussing an operator ordering issue~\cite{Weyl:1927vd}.

\section{Classical action for a non-Abelian charged particle}

A classical point particle is characterized by its position $x^\mu$
in spacetime, and if it has non-Abelian charge, also by a vector in
some internal space, corresponding to some representation of the
gauge group. For now, we will take this vector $I^a$ to be in the
Lie algebra of the gauge group, and refer to it as the `isospin' of
the particle. We can also take the internal vector to be in the
fundamental vector space of the gauge group, we will discuss that
later.

Then the following well-known pair of equations describes
the dynamics of a classical non-Abelian charged particle, first
derived from the Dirac equation by Wong \cite{Wong:1970fu},
\begin{eqnarray}
\dot{I}^a+gf_{abc}A^b_\mu I^c\dot{x}^\mu &=& 0, 
\label{Wong_parallel_transport}\\
m\frac{d}{dt}\left[\frac{\dot{x}_\mu}
{(-\dot{x}^\nu\dot{x}_\nu)^{1/2}}\right]
+g I^aF^a_{\mu\nu}\dot{x}^\nu &=& 0.
\label{eq:Wong's eqs}
\end{eqnarray}
The equation of motion for the isospin $I^a(t)$ is a parallel
transport equation along the trajectory of a particle, $x^\mu(t)$,
with the connection $\Gamma_{ac\mu}=gf_{abc}A^b_\mu$. The gauge
group will be taken to be compact semi-simple, and the structure
constants $f_{abc}$ will be chosen as real and totally
antisymmetric.  The equation of motion for $x^\mu(t)$ is a
non-Abelian generalization of the Lorentz force equation. The
parameter $t$ is a parameter along the worldline of the particle,
and a dot denotes differentiation with respect to $t$.  Let us
consider the following action,
\begin{eqnarray}
S = \int dt \left( -m(-\dot{x}^\mu\dot{x}_\mu)^{1/2}+J^a(\dot{I}^a+ 
gf_{abc}A^b_\mu \dot{x}^\mu I^c)\, \right)
\label{eq:Lagrangian1}
\end{eqnarray}
The variables appearing in this action take only real numbers as
values, and will thus correspond to real bosonic variables after
quantization.  The background metric is $\eta_{\mu\nu} =
\text{diag}(-1, 1, \cdots, 1). $ We have not included a term for
the dynamics of the gauge field, as we ignore the backreaction of
the charge on the field.  The Euler-Lagrange equation obtained by
varying $J^a(t)$ is the equation of motion for $I^a(t)$,
\begin{equation}
0=\frac{d}{dt}\frac{\partial L}{\partial\dot{J}^a} -
\frac{\partial L}{\partial J^a}
= -\dot{I}^a - gf_{abc}A^b_\mu\dot{x}^\mu I^c.
\label{eq:Parallel transport eq}
\end{equation}
Similarly, varying $I^a(t)$ gives a parallel transport equation for
$J^a(t)$,
\begin{equation}
0= \frac{d}{dt}\frac{\partial L}{\partial\dot{I}^a}-\frac{\partial
  L}{\partial I^a}= \dot{J}^a + gf_{abc}A^b_\mu \dot{x}^\mu J^c.
\end{equation}
The equation of motion for $\dot{x}^\mu(t)$ is a non-Abelian
version of the Lorentz force equation as may be expected,
\begin{equation}
m\frac{d}{dt}\left[\frac{\dot{x}_\mu}
  {(-\dot{x}^\nu\dot{x}_\nu)^{1/2}}\right]
+ gK^aF^a_{\mu\nu}\dot{x}^\nu=0,
\label{eq:Generalized Lorentz force eq}
\end{equation}
where $K^a = f_{abc}J^bI^c \equiv (J\times I)^a$. It is easy to
check that $K^a$ also satisfies the same parallel transport
equation as for $J^a$ and $I^a$. Therefore, the action
Eq.~(\ref{eq:Lagrangian1}) leads to the same set of equations
Eq.~(\ref{eq:Wong's eqs}) given by Wong.
\begin{eqnarray}
\dot{K}^a+gf_{abc}A^b_\mu\dot{x}^\mu K^c &=& 0,
\label{K-partrans}\\
m\frac{d}{dt}\left[\frac{\dot{x}_\mu}
  {(-\dot{x}^\nu\dot{x}_\nu)^{1/2}}\right]
+gK^aF^a_{\mu\nu}\dot{x}^\nu&=&0.
\end{eqnarray}
We note that the action given here is essentially the same as the
one given by Balachandran et al. \cite{Balachandran:1976ya}, but
written in terms of real variables $I$ and $J$ rather than the
complex variable $\theta^a = I^a+iJ^a$.

\subsection{Gauge invariance}

The action is invariant under the following set of 
infinitesimal transformations,
\begin{eqnarray}
\delta [I]^a &=& gf_{abc}\Lambda^bI^c,\nonumber\\
\delta [A]^a_\mu &=&
gf_{abc}\Lambda^bA^c_\mu -
\partial_\mu\Lambda^a,\nonumber\\
\Rightarrow \delta [F^a_{\mu\nu}] &=&
gf_{abc}\Lambda^bF^c_{\mu\nu},
\label{eq:gauge transformations}
\end{eqnarray}
where $ \Lambda^a $ are infinitesimal.  The same transformation
rule as for $I^a$ applies to $J^a$ and $K^a$. Under these
transformations the parallel transport equation
Eq.~(\ref{eq:Parallel transport eq}) transforms covariantly while
the generalized Lorentz force equation Eq.~(\ref{eq:Generalized
  Lorentz force eq}) is invariant. For the parallel transport
equation,
\begin{eqnarray}
\delta [\dot{I}^a + gf_{abc}(A^b_\mu\dot{x}^\mu I^c)]
&=&\frac{d}{dt}\delta[I]^a + gf_{abc}\delta[A^b_\mu]\dot{x}^\mu I^c
+ gf_{abc}A^b_\mu\dot{x}^\mu\delta[I]^c \nonumber \\
&=&gf_{akl}\dot{\Lambda}^kI^l
+gf_{akl}\Lambda^k\dot{I}^l
+gf_{abc}(gf_{bkl}\Lambda^kA^l_\mu\dot{x}^\mu-\dot{\Lambda}^b)I^c
\nonumber \\
&&
+ gf_{abc}A^b_\mu\dot{x}^\mu(gf_{ckl}\Lambda^kI^l)
\nonumber \\
&=&gf_{akl}\Lambda^k(\dot{I}^l+gf_{lbc}A^b_\mu\dot{x}^\mu I^c),
\end{eqnarray}
where we have used the Jacobi identity. Similarly,
Eq.~(\ref{eq:Generalized Lorentz force eq}) is invariant under the
gauge transformation since $F^a_{\mu\nu}$ and $K^a$ transform
covariantly.

The action has a further invariance under reparametrizations of the
worldline. If $t$ is replaced by some smooth function $\tau(t)$
with $ \dot\tau \neq 0$ , we must make the replacements
\begin{eqnarray}
  dt \to d\tau = \dot{\tau}dt\,, \qquad \frac{d}{dt} \to
  \frac{d}{d\tau} =    \frac{1}{\dot\tau}\frac{d}{dt}\,.
\end{eqnarray}
It is easy to see that the action (\ref{eq:Lagrangian1}) remains
invariant.

%

\section{Constrained Hamiltonian formalism}

The Hamiltonian plays a central role in the time evolution of
physical quantities. However, the presence of gauge symmetries
makes this a constrained system --- the variables in the Lagrangian
are not all independent. These symmetries include invariance under
the transformations of Eq.~(\ref{eq:gauge transformations}), as
well as invariance under reparametrizations of the worldline.  We
will follow the treatment due to Dirac \cite{Dirac:1967,
  Henneaux:1992ig}, which we describe briefly here.

The Lagrangian is a function of $(q,\dot{q})$. When the Hamiltonian
is obtained by the Legendre transformation, a variable $\dot{q}$ is
converted to the canonical momentum $p$ defined by
$p=\frac{\partial L}{\partial \dot{q}}$. This definition may lead
to some constraints, which are relations among the canonical
variables, of the form $\phi_m(q,p) = 0$. These are added to the
original Hamiltonian $H_0=p\dot{q}-L$ with Lagrange multipliers
$u^m(q,p)$, giving a new Hamiltonian,
\begin{equation}
 H(q,p)=H_0(q,p) + u^m(q,p)\phi_m(q,p).
\end{equation}
Next we require that the constraints should not change with time
(the consistency conditions). These conditions can lead to new
constraints as well as relations among the Lagrange multipliers.
\begin{equation}
0=\dot{\phi}_m=[\phi_m,H_0]+u^n[\phi_m,\phi_n],\label{eq:Consistency
  Condition}
\end{equation}
where the square brackets signify Poisson brackets.  That is, these
equations may give a new constraint (a secondary constraint) or
specify unknown coefficients $u^n$.

(a) If a new constraint is found, it is again applied to the
consistency condition in order to see whether it gives a new
further constraint or restricts the coefficients $u^n$. This
algorithm ends if no more new constraint comes out.

(b) If relations are found among the Lagrange multipliers, $u^n$,
general solutions for $u^n$ to Eq.~(\ref{eq:Consistency Condition})
are written in the form $u^n=U^n+V^n$, where $U^n$ are particular
solutions and hence specified, and $V^n$ are homogeneous solutions,
which are in a form of linear combination of the solutions with
arbitrary coefficients $v^a$, $V^n=v^aV^n_a$.

Any constraint which has vanishing Poisson brackets with all
constraints is called a first class constraint and all other
constraints are called second class constraints. The consistency
conditions ensure that the Hamiltonian has vanishing Poisson
brackets with all the constraints. Ultimately all the constraints
including second class constraints need to be imposed on the system
in order to keep only independent variables in the theory.

In order to do this, we first define Dirac brackets using all the
second class constraints, denoted by $\chi_\alpha,$
\begin{equation}
[F,G]^* \equiv [F,G] - [F,\chi_\alpha]C^{\alpha\beta}[\chi_\beta,G],
\label{eq:Definition of the Dirac bracket}
\end{equation}
where $C$ is the matrix given by the matrix elements
$C_{\alpha\beta}=[\chi_\alpha,\chi_\beta]$ and the inverse elements
$C^{\alpha\beta}\equiv[C^{-1}]_{\alpha\beta}$. The Dirac bracket,
like the Poisson bracket, is antisymmetric and satisfies the Jacobi
identity, but the Dirac bracket of any quantity with a second class
constraint is zero,
\begin{equation}
[\chi_\alpha,F]^*=0.
\end{equation}
This implies that by replacing the Poisson brackets by Dirac
brackets $[F,G]^*$, we can set all constraints, including the
second class constraints $\chi_\alpha$, equal to zero at the end of
a calculation.

\subsection{Worldline formalism}

For technical convenience we will use the worldline formalism to
write the action \cite{Feynman:1950ir, Kleinert:2006ev}, and then
to compute the path integral, for the particle with non-Abelian
charge. The action in Eq.~(\ref{eq:Lagrangian1}) can be rewritten
in the worldline formalism by introducing a Lagrange multiplier
$h$,
\begin{equation}
S=\int dt(L_0+L_I),\label{worldline action}
\end{equation}
where
\begin{equation}
L_0=\frac{1}{2}h^{-1}\dot{x}^2 - h\frac{m^2}{2},
\qquad L_I=J^a[\dot{I}^a + g(A_\mu\dot{x}^\mu\times I)^a].
\label{worldline lagrangian}
\end{equation}
The introduction of $h$ makes the action in the worldline formalism
invariant under the reparametrizations
\begin{equation}
t\rightarrow f(t),\qquad h\rightarrow h/(df/dt).
\end{equation}
The equation of motion for $h$ is
\begin{equation}
\dot{x}^2 + h^2m^2=0,
\label{eq:h-eom}
\end{equation}
while the other equations of motion are now
\begin{eqnarray}
\dot{I}^a+gf_{abc}A^b_\mu I^c\dot{x}^\mu &=& 0, 
\label{I-eom}\\
\frac{d}{dt}\left(h^{-1}\dot{x}_\mu\right)
+ gK^aF^a_{\mu\nu}\dot{x}^\nu &=& 0,
\label{eq:x-I-eom}
\end{eqnarray}
where $K^a = f_{abc}J^bI^c\equiv (J\times I)^a$ as before.
Using Eq.~(\ref{eq:h-eom}) in Eq.~(\ref{eq:x-I-eom}), we recover Eq.~(\ref{eq:Wong's eqs}). Alternatively, if
Eq.~(\ref{eq:h-eom}) is plugged back into the action, the previous
Lagrangian Eq.~(\ref{eq:Lagrangian1}) is recovered. The canonical
momenta can be calculated from Eq.~(\ref{worldline lagrangian}),
\begin{eqnarray}
P^a_J&=&0,\\
P^a_I&=&J^a,\\
P_h&=&0,\\
P_\mu&=&h^{-1}\dot{x}_\mu-gA^a_\mu K^a,
\end{eqnarray}
from which we can read off the constraints immediately,
\begin{eqnarray}
\phi^a_1&=&P^a_J \approx 0,\nonumber\\
\phi^a_2&=&P^a_I-J^a \approx 0,\nonumber\\
\phi_h&=&P_h \approx 0,
\label{eq:Constraints in Worldline Formalism}
\end{eqnarray}
where the symbol $\approx$ indicated weak equality, i.e., equality
on the submanifold defined by constraints.  Now we write down the
canonical Hamiltonian $H_0$,
\begin{eqnarray}
H_0&=&P_\mu\dot{x}^\mu + P^a_I\dot{I}^a + P_J\dot{J}^a +
P_h\dot{h}-(L_0+L_I)\nonumber\\
&=&\frac{h}{2}[(P_\mu+gA^a_\mu K^a)^2+m^2]  + P^a_J\dot{J}^a +
(P^a_I-J^a)\dot{I}^a + P_h\dot{h},
\label{canonical H_0}
\end{eqnarray}
where it has been assumed that the time derivatives of the
coordinates, e.g. $\dot{I}$, are expressed in terms of the
canonical variables, in principle. If the constraints
Eq.~(\ref{eq:Constraints in Worldline Formalism}) are added to
$H_0$, the new Hamiltonian $H_1$ remains in the same form as
before,
\begin{eqnarray}
H_1&=&H_0+ C^a_1 P^a_J + C^a_2 (P^a_I-J^a) + C_hP_h \nonumber \\
&=&\frac{h}{2}[(P_\mu + gA^a_\mu K^a)^2+m^2] +
P^a_J[C^a_1 +\dot{J}^a(p,q)]+(P^a_I-J^a)[C^a_2 +\dot{I}^a(p,q)] \nonumber \\
&&+P_h[C_h+\dot{h}(p,q)].
\label{canonical H_1}
\end{eqnarray}
The consistency conditions require that the constraints have
vanishing Poisson brackets with $H_1$, so writing $\Pi_\mu = (P_\mu
+ gA^a_\mu K^a)\,,$ we find
\begin{eqnarray}
0=\dot{\phi}^a_1 &=& [\phi^a_1 , H_1]\nonumber \\
&=&[C^a_2 + \dot{I}^a(p,q)] + 2g(f_{abc}A^b_\mu
I^c)(h/2)\Pi^\mu, \label{phi1dot} \\
0=\dot{\phi}^a_2 &=& [\phi^a_2 , H_1] \nonumber \\
&=&-[C^a_1 + \dot{J}^a(p,q)] - 2g(f_{abc}A^b_\mu
J^c)(h/2)\Pi^\mu, \label{phi2dot} \\
0=\dot{\phi}_h&=&[P_h,H_1]\nonumber \\
&=&-\frac{1}{2}[(P_\mu+gA^a_\mu K^a)^2+m^2].\label{phi_hdot}
\end{eqnarray}
The first two relations allow us to eliminate $C^a_2 +
\dot{I}^a(p,q)$ and $C^a_1 + \dot{J}^a(p,q)$ in $H_1$. The last
equation is a secondary constraint, which we denote by $\phi_3$\,,
\begin{equation}
\phi_3=(P_\mu+gA^a_\mu K^a)^2+m^2 \approx 0.
\end{equation}
Now we can rewrite $H_1$ with the unknown coefficients eliminated,
\begin{eqnarray}
H_1&=&\frac{h}{2}(\Pi^\mu\Pi_\mu+m^2)+(C_h+\dot{h})P_h \nonumber \\
&&+\frac{h}{2}[- 2g(f_{abc}A^b_\mu J^c) \Pi^\mu P^a_J
- 2g(f_{abc}A^b_\mu I^c) \Pi^\mu (P^a_I-J^a) ],
\end{eqnarray}
where we have defined $\Pi_\mu=P_\mu+gA^a_\mu K^a$.
We can see that $\phi_3$ identically commutes with $H_1$ without
giving any further constraint or condition for the coefficients,
\begin{eqnarray}
\dot{\phi}_3&=&[\phi_3,H_1]\,\nonumber \\
&=&\frac{h}{2}\left(- 2gf_{abc}A^b_\mu J^c \Pi^\mu
  [\Pi^2+m^2,P^a_J] - 2gf_{abc}A^b_\mu I^c \Pi^\mu
  [\Pi^2+m^2,P^a_I-J^a] \right)\,\nonumber
\\
&=&\frac{h}{2}\left(- 4gf_{abc}A^b_\mu J^c
  \Pi^\mu\Pi^\nu[\Pi_\nu,P^a_J] - 4gf_{abc}A^b_\mu I^c \Pi^\mu
  \Pi^\nu [\Pi_\nu,P^a_I] \right) = 0.
\label{phi3dot}
\end{eqnarray}
Also, by the following consistency condition, $C_h$ must vanish,
\begin{equation}
\dot{h}=[h,H_1]=(C_h+\dot{h})[h,P_h] = C_h + \dot{h}.
\end{equation}
Thus the final expression for the Hamiltonian is
\begin{eqnarray}
H &=& h\left[\frac12(\Pi^\mu\Pi_\mu+m^2) - g f_{abc}A^b_\mu
  J^c P^a_J  - g f_{abc}A^b_\mu I^c (P^a_I-J^a)\right]
+ \dot{h}P_h\,,
\label{eq:Final H}
\end{eqnarray}
with the constraints
\begin{eqnarray}
\phi^a_1 &=& P^a_J \approx 0,\\
\phi^a_2 &=& P^a_I-J^a \approx 0,\\
\phi_3 &=& (P_\mu + gA^a_\mu K^a)^2+m^2 \approx
0, \label{phi3.constraint}\\
\phi_h &=& P_h \approx 0.
\end{eqnarray}
Note that $\phi^a_1, \phi^a_2$ and $\phi_3$ can be recombined into
one first class and two second class constraints. The term within
the square brackets in Eq.~(\ref{eq:Final H}) is the first class
linear combination of these constraints. We may call this term
$\phi_0$. Then we can choose $\phi^a_1$ and $\phi^a_2$ as the
independent second class constraints, and $\phi_0$ and $\phi_h$ as
the independent first class constraints. In other words, the
constraint in Eq.~(\ref{phi3.constraint}) is replaced by
$\phi_0\approx 0$. We note that the Hamiltonian of
Eq.~(\ref{eq:Final H}) is also a first class constraint by
construction. If the gauge freedom in terms of arbitrary
coefficients is considered, $H$ does not explicitly show the gauge
freedom for the first class secondary constraint $\phi_0$. For
this, one can use the extended Hamiltonian, where the number of
unknown coefficients explicitly matches with that of the first
class constraints,
\begin{eqnarray}
H_E=H+\frac{C_0}{2}\phi_0.
\end{eqnarray}
Recognizing $\dot{h}$ is arbitrary we see that the number of the
arbitrary coefficients is equal to that of the first class
constraints.

We can calculate the Dirac brackets using the matrix
\begin{eqnarray}
C_{12}^{ab} = \delta_{ab} = - C_{21}^{ab},
\label{Eq:PBmatrix-C12}
\end{eqnarray}
and set the second class constraints to zero due to the property
$[\chi_\alpha, F]^*=0$ where $\chi_\alpha$ is the second class
constraint and $F$ is an arbitrary quantity. The constraint
$P_h\approx0$, along with the gauge fixing condition $h=\lambda$,
which we now set, allows us to remove a pair of variables $(h,P_h)$
\cite{Kleinert:2006ev}. Then we find
\begin{equation}
H_E = C(\Pi^\mu\Pi_\mu+m^2),
\end{equation}
as also given in~\cite{Balachandran:1976ya}.  One can see whether
we use $H$ or $H_E$ the path integral will be in the same form as
long as all the corresponding gauge conditions to the first
constraints are applied for it.

\section{Path integral}\label{path integral}

A general form of the path integral for a constrained system is
given in \cite{Senjanovic:1976br, Henneaux:1992ig}.
\begin{equation}
  \int\mathscr{D}z^A\prod_{t,\alpha}\delta(\chi_\alpha)
  (\det[\chi_\alpha,\chi_\beta])^{1/2}
  \prod_{t,a}\delta(\phi_a)\delta(G_a)
  \prod_{t}\det[G_a,\phi_b]e^
  {\frac{i}{\hbar}S[z^a(t)]\label{eq:General Path Integral}}, 
\end{equation}
where $\phi_a$ and $\chi_\alpha$ are a first and a second class
constraints, respectively and $G_a$ is a gauge fixing condition for
$\phi_a$. Here $\delta(\chi_\alpha)$ is responsible for the second
class constraints and $(\det[\chi_\alpha,\chi_\beta])^{1/2}$ for
their Jacobian factor. After suitable gauge conditions for the
first class constraints are chosen, a gauge condition and its
corresponding first class constraint form a pair of second class
constraints. Thus the delta functions corresponding to the first
class constraint and the gauge fixing condition, $\delta(\phi_a)$
and $\delta(G_a)$, respectively, are similarly incorporated
together with the Jacobian factor $\det[G_a,\phi_b]$ in the path
integral. The gauge fixing constraints need to be introduced in
order to fix the arbitrary coefficients on the corresponding first
class constraints in the Hamiltonian. Applying the formula
Eq.~(\ref{eq:General Path Integral}) to our case, we find that the
path integral becomes
\begin{equation}
\int^\infty_0
d\lambda\int\mathscr{D}x^\mu\mathscr{D}P_\mu
\mathscr{D}I^a\mathscr{D}P^a_I\mathscr{D}h\delta[h-\lambda]
[\phi_0,\chi_0]\delta[\phi_0]e^{\frac{i}{\hbar}\int
  dt\mathscr{L}_{in}|_{P^a_J=0,J^a=P^a_I}}.
\label{eq:Path Integral Lin} 
\end{equation}
The Jacobian factor $(\det[\chi_\alpha,\chi_\beta])^{1/2}=1$ and
the delta functions for the second class constraints have been used
by setting $P^a_J$ to zero and replacing $J^a$ by $P^a_I$. The
integration over the gauge choice $\lambda$ for the first class
constraint $P_h\approx0$ implies that the gauge freedom, i.e., a
reparametrization invariance, is factored out
\cite{Kleinert:2006ev}. $\mathscr{L}_{in}$ is
\begin{eqnarray}
\mathscr
{L}_{in}&=&P_\mu\dot{x}^\mu+P^a_I\dot{I}^a+P^a_J\dot{J}^a+P_h\dot{h}-H
\nonumber\\ 
&=&\Pi_\mu\dot{x}^\mu-gA^a_\mu\dot{x}^\mu
K^a+P^a_I\dot{I}^a-h\phi_0.
\end{eqnarray}
We already noted that the Hamiltonian itself is a constraint and
hence zero. $\phi_0$ can also be placed in the same footing. One
can express the delta function for $\phi_0$ in the following form
\begin{equation}
\delta[\phi_{0}]=\frac{\hbar}{2\pi}\int \mathscr{D}\omega
e^{\frac{i}{\hbar}\int dt\omega\phi_0}. 
\end{equation}
Hence, the delta function for $\phi_0$ can be combined into the
Hamiltonian by redefining $h-\omega\rightarrow h$, 
\begin{eqnarray}
\mathscr{D}h\delta[\phi_0]e^{\frac{i}{\hbar}\int \mathscr{L}_{in}(h)dt}
&\sim&\int \mathscr{D}h\mathscr{D}\omega e^{\frac{i}{\hbar}\int
  \mathscr{L}_{in}(\omega-h)dt}\nonumber\\ 
&=&\int \mathscr{D}h\mathscr{D}\omega e^{\frac{i}{\hbar}\int
  \mathscr{L}_{in}(h)dt}\nonumber\\ 
&\sim&\int \mathscr{D}he^{\frac{i}{\hbar}\int \mathscr{L}_{in}(h)dt}
\end{eqnarray}
Therefore, including the delta function for the constraint $\phi_0$
is redundant. 

After the redefinition of $h$ we put the gauge condition
$\delta[h-\lambda]$. Then we notice that $\mathscr {L}_{in}$
becomes different from the original Lagrangian $L_0+L_I$,
Eq.~(\ref{worldline lagrangian}), only by
$-\frac{h}{2}\left(\Pi-\frac{1}{h}\dot{x}\right)^2$.
\begin{eqnarray}
\mathscr
{L}_{in}&=&P_\mu\dot{x}^\mu+P^a_I\dot{I}^a+P^a_J\dot{J}^a+P_h
\dot{h}-H\nonumber\\ 
&=&\Pi_\mu\dot{x}^\mu-gA^a_\mu\dot{x}^\mu
K^a+P^a_I\dot{I}^a-\frac{h}{2}\Pi^2-\frac{h}{2} m^2\nonumber\\
&=&-\frac{h}{2}\left(\Pi-\frac{1}{h}\dot{x}\right)^2
+\frac{1}{2h}\dot{x}^2-\frac{h}{2}m^2
+J^a[\dot{I}^a+g(A_\mu\dot{x}^\mu\times I)^a]\nonumber\\
&&+(P^a_I-J^a)\dot{I}^a+P^a_J\dot{J}^a.
\end{eqnarray}
where $\dot{x}^\mu\equiv\frac{x_{n+1}-x_{n}}{\Delta t}$,etc, so
$\mathscr {L}_{in}$ has been expressed only by canonical variables
$(p,q)$, i.e., $\mathscr {L}_{in}=\mathscr{L}_{in}(p,q)$. After the
terms $(P^a_I-J^a)\dot{I}^a$ and $P^a_J\dot{J}^a$ are removed by
the second class constraints with the Dirac bracket formalism, the
Lagrangian becomes as follows.
\begin{equation}
\mathscr
{L}_{in}\Rightarrow-\frac{h}{2}\left(\Pi-\frac{1}{h}
  \dot{x}\right)^2 
+\frac{1}{2h}\dot{x}^2-\frac{h}{2}m^2
+P^a_I[\dot{I}^a+g(A_\mu\dot{x}^\mu\times I)^a].\label{eq:Lin}
\end{equation}
According to the definition of the momentum $P_\mu$, the first term
would be zero in continuum limit. However, one cannot simply set it
equal to zero since $\dot{x}^\mu=\frac{x_{n+1}-x_{n}}{\Delta t}$ is
treated as independent variables in the path integral. This implies
that the path integral directly expressed from the given Lagrangian
may be different from one reconstructed from the Hamiltonian. Our
interest in this paper is in the integration over the internal
variables $(P^a_I,I^a)$. In the next subsection we see that a
certain gauge choice can allow us to integrate over the internal
degrees of freedom.

\subsection{Gauge condition for $\phi_0$} \label{sec:Gauge
  condition for phi0} 
Here we choose a gauge fixing condition $\chi_0$ for $\phi_0$. The
gauge condition comes into the path integral as the delta function
$\delta[\chi_0]$ with the Jacobian $[\phi_0,\chi_0]$. We are
interested in the case that the gauge fixing does not spoil form of
the isospin part so that later we are able to integrate over the
internal variables separately. We choose the gauge fixing
condition,
\begin{equation}
\chi_{0}=x^{0}-t=0.
\end{equation}
It is easy to check the Poisson bracket $[\phi_0,\chi_0]$ is non-zero.
\begin{eqnarray}
[\phi_0,\chi_0]&=&\left[\frac{1}{2}(\Pi^2+m^2)+C^a_1\phi^a_1 +
  C^a_2\phi^a_2,x^{0}-t\right]\nonumber\\  
&=&\left[\frac{1}{2}(\Pi^2+m^2),x^{0}-t\right]\nonumber\\
&=&\frac{1}{2}\times2\Pi^\mu[P_\mu+gA^a_\mu K^a,x^{0}]\nonumber\\
&=&-\Pi^{0}.
\end{eqnarray}
By a change of variables $\Pi_\mu-\frac{1}{h}\dot{x}_\mu\rightarrow
P_\mu$, the Jacobian factor $[\phi_0,\chi_0]$ becomes
$\left(-P^{0}-\frac{1}{h}\dot{x}^0\right)$. The relevant path
integral can then be written as
\begin{eqnarray}
&&\int \mathscr{D}h\delta[h-\lambda]\mathscr{D}x^{0}
\delta[x^{0}-t]\mathscr{D}x^i\mathscr{D}P^ie^{\frac{i}{\hbar}\int
  dt\left\{-\frac{h}{2}P_i^2 
+\frac{1}{2h}\dot{x}_i^2-\frac{h}{2}m^2
+P^a_I[\dot{I}^a+g(A_\mu\dot{x}^\mu\times I)^a]
\right\}}\nonumber\\ 
&&\times\mathscr{D}P^{0}\left[-P^{0}-\frac{1}{h}
  \dot{x}^{0}\right]e^{\frac{i}{\hbar}\int 
  dt\left[-\frac{h}{2}(P_0)^2 
+\frac{1}{2h}(\dot{x}_0)^2\right]}\nonumber\\
 &\sim&-\int
 \mathscr{D}h\delta[h-\lambda]\mathscr{D}x^ie^{\frac{i}{\hbar}\int
   dt\left\{\frac{1}{2h}\dot{x}_\mu^2-\frac{h}{2}m^2 
+P^a_I[\dot{I}^a+g(A_\mu\dot{x}^\mu\times I)^a]
 \right\}_{x^0=t}},
\end{eqnarray}
where the Gaussian integrals over $P^\mu$ have been taken.

\subsection{Integration over internal variables}

In the last subsection we have chosen the gauge fixing condition
for $\phi_0$. Since this gauge fixing does not change structure of
the isospin part, we are able to compute an integration over
isospin variables. Taking the relevant part of the action in a
time-sliced form,
\begin{eqnarray}
S_I&=&\displaystyle\sum^{N+1}_{n=1}
P^a_{I,n}[I^a_n-I^a_{n-1}+gf_{abc}A^b_{\mu,n} (x^\mu_{n}-x^\mu_{n-1})
I^c_n]\nonumber\\
&=&\displaystyle\sum^{N+1}_{n=1} P^a_{I,n}(M^{ac}_nI^c_{n}-I^c_{n-1}),
\end{eqnarray}
where $M^{ac}_n\equiv\delta^{ac}+gf_{abc}A^b_{\mu,n}
(x^\mu_{n}-x^\mu_{n-1})$. The integral over $P^a_{I,n}$ in the path
integral produces a delta function,
\begin{equation}
\int dP^a_{I,n}e^{\frac{i}{\hbar}P^a_{I,n}(M^{ac}_nI^c_n-I^c_{n-1})}
=2\pi\hbar\delta(M^{ac}_nI^c_{n}-I^c_{n-1}).
\end{equation}
What finally remains in the path integral is a product of all the
delta functions and after integration over $I^a$ it reduces to a
single delta function,
\begin{eqnarray}
&&(2\pi\hbar)^{N+1}\int dI^a_N\cdots dI^a_1
\delta(M_{N+1}I_{N+1}-I_N)\delta(M_{N}I_{N}-I_{N-1})
\cdots\delta(M_{1}I_{1}-I_{0})\nonumber\\
&=&(2\pi\hbar)^{N+1}\int dI^a_{N-1}\cdots
dI^a_1\delta(M_{N}M_{N+1}I_{N+1}-I_{N-1})
\cdots\delta(M_{1}I_{1}-I_0)\nonumber\\
&=&(2\pi\hbar)^{N+1}\delta(M_{1}\cdots
M_{N}M_{N+1}I_{N+1}-I_{0}).
\end{eqnarray}
The matrix product $M_{1}\cdots M_{N}M_{N+1}$ is a finite parallel
transport operator backward in time, but it can be placed in
time order. Let us define $(A_n)_{ac}=gf_{abc}A^b_{\mu,n}
(x^\mu_{n}-x^\mu_{n-1})/\Delta t_{n}$, where $\Delta
t_{n}=t_n-t_{n-1}$. Then using the antisymmetry of $A_n$,
we can write
\begin{eqnarray}
&&[M_{1}\cdots M_{N}M_{N+1}]_{ab}I^b_{N+1}\nonumber\\
&=&[(1+A_1\Delta
t_1)\cdots(1+A_{n+1}\Delta t_{n+1})\cdots(1+A_{N+1}\Delta
t_{N+1})]_{ab}I^b_{N+1}\nonumber\\
&=&I^b_{N+1}\left[(1-A_{N+1}\Delta t_{N+1})\cdots(1-A_{n+1}\Delta
t_{n+1})\cdots(1-A_1\Delta t_1)\right]_{ba}\nonumber\\
&=&\left[I_{N+1}Te^{-\int^{t_{N+1}}_{t_0} Adt}\right]^a,
\end{eqnarray}
 The path integral for the isospin part thus becomes
\begin{eqnarray}
&&\int dI^a_N\cdots dI^a_1
\delta(M_{N+1}I_{N+1}-I_N)\delta(M_{N}I_{N}-I_{N-1})
\cdots\delta(M_{1}I_{1}-I_{0})\nonumber\\
&\propto&\delta(I_fTe^{-\int^{t_f}_{t_i}
  Adt}-I_i).\label{eq:delta function constraint}
\end{eqnarray}
We know that the solution of the classical equation
$\dot{I}^a+gf_{abc}A^b_\mu\dot{x}^\mu=0$ is
\begin{equation}
I_f=Te^{-\int^{t_f}_{t_i} A dt}I_i. \label{eq:Solution of parallel
  transport eq}
\end{equation}
The unitary matrix $U=Te^{-\int^{t_f}_{t_i} A dt}$, which consists
of real elements, has an inverse element
$[U^{-1}]_{ab}=[U^\dagger]_{ab}=U_{ba}$. Thus, it can be seen that
the solution of the parallel transport equation,
Eq.~(\ref{eq:Solution of parallel transport eq}) is equivalent to
Eq.~(\ref{eq:delta function constraint}), that is,
\begin{eqnarray}
&&I_f=UI_i\nonumber\\
&\Rightarrow& [U^{-1} I_f]^a=[I_i]^a\nonumber\\
&\Rightarrow& U_{ba}[I_f]^b=[I_fTe^{-\int^{t_f}_{t_i} A dt}]^a=[I_i]^a.
\end{eqnarray}
We see that the path integral agrees with the classical result.  As
a result it contributes as a constraint, the solution to the
classical parallel transport equation. The path integral can be
written in the form before the integration over the initial
momentum $P^a_{I}(t_i)\equiv j^a$
\begin{equation}
\int^\infty_{-\infty}dj\int^\infty_0d\lambda
\int\mathscr {D}h\delta(h-\lambda)\int
\mathscr{D}xe^{\frac{i}{\hbar}\int
  dt[\frac{1}{2h}\dot{x}^2-\frac{h}{2}m^2]}
e^{\frac{i}{\hbar}j(I_fTe^{-\int^{t_f}_{t_i}  
  Adt}-I_i)},\\\label{eq:final path integral}
\end{equation}
where $A_{ac}\equiv gf_{abc}A^b_\mu\dot{x}^\mu$. Note that here we
did not apply the gauge fixing for $\phi_0$ from the subsection
\ref{sec:Gauge condition for phi0} because we will derive the
Klein-Gordon constrained wave equation which has a gauge symmetry.
In the next section and in Appendix \ref{Appx:Derivation of the
  Constraint Wave Equation}, the validity of this path integral
will be supported by deriving the wave equation obtained as a
constrained wave equation in the canonical quantization.  In
addition, Appendix \ref{Appx:Derivation of the Equation of Motion}
provides the derivation of the generalized Lorentz force equation
Eq.~(\ref{eq:Generalized Lorentz force eq}) by varying the path
integral.

\section{ Fundamental representation} 
\label{sec:Fundamental Representation}

So far we have chosen the internal degrees of freedom of the
classical particle to be in the adjoint representation. It is not
difficult to write an action with the internal degrees in the
fundamental representation. In the action of
Eq.~(\ref{eq:Lagrangian1}) we first write the structure constants
$f_{abc}$ as $-i(T^b)_{ac}$ where $T^b$ are the generators.  Then
the `charge vectors' $I$ and $J$ are taken in the vector space of
the fundamental representation, so that we can replace
$f_{abc}A^bJ^aI^c\rightarrow -iA^bT^b_{ij}J^iI^j$. If the
fundamental representation is real, for example if the gauge group
is SO(N), we can choose the matrices $T^a_{ij}$ to be antisymmetric
and purely imaginary. Then the Lagrangian becomes
\begin{equation}
L=\frac{1}{2h}\dot{x}^2-
\frac{h}{2}m^2 + J^i(\dot{I}^i-igT^a_{ij}A^a_\mu I^j\dot{x}^\mu).
\label{real-fundamental}
\end{equation}
If the fundamental vector space is complex, for example when the
gauge group is SU(N), the vectors $I$ and $J$ will be complex. Then 
the Lagrangian of Eq.~(\ref{real-fundamental}) doubles the internal
degrees of freedom. In order to avoid this doubling, we replace
in this Lagrangian
\begin{eqnarray}
&&J^a\dot{I}^a\to iI^{i*}\dot{I}^i\nonumber\\
&&J^af_{abc}A^b_\mu I^c\to iI^{i*}(-iT^b_{ij})A^b_\mu I^j,
\end{eqnarray}
where now the $T^a$ are Hermitian, but not all purely imaginary.
Then we can write the Lagrangian as
\begin{equation}
L=\frac{1}{2h}\dot{x}^2-
\frac{h}{2}m^2+iI^{*i}(\dot{I}^i-igT^a_{ij}A^a_\mu
I^j\dot{x}^\mu). 
\label{complex-fundamental}
\end{equation}
We can then write the parallel transport equation 
for the complex charge vector $I$, 
\begin{eqnarray}
\dot{I}^i-igT^a_{ij}A^a_\mu I^j\dot{x}^\mu&=&0, 
\label{complex-fund-transport-eq}
\end{eqnarray}
and the generalization of the Lorentz force equation is now
\begin{eqnarray}
\frac{d}{dt}\left(\frac{m\dot{x}_\mu}{\sqrt{-\dot{x}^2}}\right)
&+&\frac{d}{dt}(gA^a_\mu T^a_{ij}I^{*i}I^j)
- gT^a_{ij}I^{*i}I^j(\partial_\mu
A^a_\nu)\dot{x}^\nu \nonumber\\
&=& \frac{d}{dt}\left(h^{-1} \dot{x}_\mu\right)
+gK^aF^a_{\mu\nu}\dot{x}^\nu 
= 0\,,
\end{eqnarray}
where now we have defined $K^a = -T^a_{ij}I^{*i}I^j = K^{*a}$. 
The Lorentz force equation 
thus has the same form as Eq.~(\ref{eq:x-I-eom}), i.e. the same form for 
adjoint and fundamental representations.
It can be easily seen that
$K^a$ satisfies the same parallel transport equation as in
Eq.~(\ref{K-partrans}),
\begin{eqnarray}
\dot{K}^a&=&-T^a_{ij}(\dot{I}^{*i}I^j+I^{*i}\dot{I}^j)\nonumber\\
&=&igI^{*i}I^j(T^bT^a-T^aT^b)_{ij}A^b_\mu \dot{x}^\mu\nonumber\\
&=&-gf_{abc}A^b_\mu K^c\dot{x}^\mu.
\end{eqnarray}
The constraints are similar to those in the adjoint
representation,
%
%
%
\begin{eqnarray}
\phi^i_1&=&P^i_{I^*} \approx 0,\\
\phi^i_2&=&P^i_I-iI^{*i}\approx 0,\\
\phi_h&=&P_h \approx 0.
\end{eqnarray}
We also note that 
\begin{eqnarray}
P_\mu&=&h^{-1}\dot{x}_\mu-gA^a_\mu K^a\,.
\end{eqnarray}
After removing the variables $I^{*i}$ and $P^i_{I^*}$, similarly to
Eq.~(\ref{eq:Lin}), the Lagrangian in the fundamental
representation, $\mathscr{L}^f_{in}$, can be written down,
\begin{eqnarray}
\mathscr
{L}^f_{in}&=&P_\mu\dot{x}^\mu+P^i_I\dot{I}^a+P^i_{I^*}
\dot{I}^{*i}+P_h\dot{h}-H\nonumber\\  
&=&-\frac{h}{2}\left(\Pi-\frac{1}{h}\dot{x}\right)^2
+\frac{1}{2h}\dot{x}^2-\frac{h}{2}m^2
+P^i_I\left(\dot{I}^i-igT^b_{ij}A^b_\mu I^j\dot{x}^\mu\right).
\end{eqnarray}
And the path integral becomes
\begin{equation}
\int^\infty_0
d\lambda\int\mathscr{D}x^\mu\mathscr{D}P_\mu
\mathscr{D}I^i\mathscr{D}P^i_I\mathscr{D}h\delta[h-\lambda]
[\phi_0,\chi_0]\delta[\phi_0]e^{\frac{i}{\hbar}\int
  dt\mathscr{L}^f_{in}}. \label{eq:Path Integral Lin fundamental}
\end{equation}

\section{Derivation of the constrained wave equation} 
\label{sec:Derivation of Wave equation}

In this section we derive the wave equation satisfied by the path
integral. This links the classical formulation with the quantum
one, in turn justifying the choice of the classical action. While
we expect to find some generalization of the Klein-Gordon equation,
the operator representation from the classical Hamiltonian brings
ambiguity in ordering of operators \cite{Mizrahi:1975pw}. We will
see that the quantum Hamiltonian operator as derived from our path
integral includes a term which does not remain invariant under the
gauge transformations of Eq.~(\ref{eq:gauge transformations}). We
will verify this result by taking an inverse Weyl transformation of
the classical Hamiltonian. The derivation is long, so for the sake
of clarity we have gathered some of the intermediate calculations
in Appendix \ref{Appx:Derivation of the Constraint Wave Equation}.

The wave function at the final point $(x_f, I_f, t_f)$ is the 
weighted sum of the
wave functions at all possible starting positions
$(x_i, I_i, t_i)\,,$ weighted by the kernel
\begin{equation}
K(x_f, I_f, t_f; x_i, I_i, t_i)	 \equiv
\langle q_f, t_f | q_i, t_i \rangle\,,
\end{equation}
where $K(x_f, I_f, t_f; x_i, I_i, t_i)$ is the path integral
Eq.~(\ref{eq:final path integral}) and $q$ stands for all
variables, both internal and external. We are interested in the
differential equation satisfied by the wave function. So we
consider an infinitesimal evolution with $t_i=t$ and $t_f=t+\Delta
t$. We also write $x_f=x$, $I_f=I$ and $x_i=x-\xi$,
$I_i=I-\eta$. Then we can write the wave function as
\begin{equation}
\psi(x,t+\Delta t)= N\int d\eta\int d\xi\psi(x-\xi,I-\eta,t)
K(x, I, t + \Delta t; x - \xi, I - \eta, t),
\label{Eq:wavefunction}
\end{equation}
where $N$ is a proportionality constant to be determined by
matching wave functions in different times. Since $\Delta t$ is
infinitesimal, we can write
\begin{eqnarray}
\langle q_f, t+\Delta t | q_i, t \rangle &=& \langle q_f |
e^{-\frac{i}{\hbar}\hat{H}\Delta t} | q_i  \rangle \nonumber \\
&\sim& \int dp\, e^{i(q_f - q_i)p/\hbar}
e^{-\frac{i}{\hbar}H_c(p,(q_f+q_i)/2)\Delta t}
\nonumber \\
&\sim& e^{\frac{i}{\hbar}\mathscr{L}\Delta t},
\end{eqnarray}
where $\hat{H}$ is a Hamiltonian operator, $H_c(p,q)$ is a
Hamiltonian function and $\mathscr{L}$ is a Lagrangian in a
discrete time interval. Here $p$ stands for the momenta canonically
conjugate to all variables. We identify the Lagrangian
$\mathscr{L}$ from Eq.~(\ref{eq:final path integral}) and expand
$Te^{-\int^{t_f}_{t_i} Adt^\prime}$ to first order in
$\epsilon\equiv\lambda\Delta t$, using
$[A]_{ac}=gf_{abc}A^b_\mu\dot{x}^\mu$ to write
\begin{eqnarray}
\mathscr{L}\Delta
t&=& \left(\frac{1}{2\lambda}\dot{x}^2-\lambda\frac{m^2}{2}\right)
\Delta t +j\left(I_fTe^{-\int^{t_f}_{t_i} Adt^\prime}-I_i\right)
\nonumber \\ 
&=& \frac{1}{2\epsilon}(x_f-x_i)^2-\frac{m^2}{2}\epsilon
+j^a\left[I^b_f(\delta_{ba}-gf_{bca}A^c_\mu(x_f-x_i)^\mu)
  -I^a_i\right] + {\cal O}(\epsilon^2)\nonumber \\
&=&\frac{1}{2\epsilon}\xi^2-\frac{m^2}{2}\epsilon
+j^a\left(\eta^a + gf_{acb}A^c_\mu\xi^\mu I^b_f\right),
\label{eq:Ldelta-t}
\end{eqnarray}
where $\eta = I_f - I_i$ and $\xi = (x_f - x_i)\,$. According to
the Weyl correspondence~\cite{Mizrahi:1975pw} the classical
Hamiltonian $H_c$ has to be evaluated at the midpoint
$\frac{q_{j+1}+q_j}{2}$,
\begin{eqnarray}
\langle q_{j+1}|e^{-i(t_{j+1}-t_j)\hat{H}/\hbar}|q_j\rangle
&\propto& \langle q_{j+1}|\int
e^{-i(t_{j+1}-t_j)H_c(p,q)/\hbar}\Delta(p,q)dpdq|q_j \rangle
\nonumber \\
&\propto&\int dp
e^{\frac{i}{\hbar}\left[p\cdot\left(\frac{q_{j+1} -
	q_j}{t_{j+1}-t_j}\right)  -
    H_c\left(p,\frac{q_{j+1}+q_j}{2}\right)\right](t_{j+1}-t_j)},
\end{eqnarray}
with
\begin{equation}
\Delta(p,q)=\int due^{iq\cdot u/\hbar}|p-u/2\rangle\langle p+u/2|.
\end{equation}
Thus the isospin dependent part of the action takes the midpoint
value,  denoted by a `bar', 
\begin{equation}
S_I=P^a_{I,n}[I^a_n-I^a_{n-1}+gf_{abc}
\overline{A^b_{\mu,n}I^c_n}(x^\mu_n-x^\mu_{n-1})].
\end{equation}
Using the expression for the path
integral from Eq.~(\ref{eq:final path integral}) in 
Eq.~(\ref{Eq:wavefunction})
and using Eq.~(\ref{eq:Ldelta-t}), we can write the
wave function at $t+\Delta t$ as
\begin{equation}
\psi(x,t+\Delta t)=N\int d\eta\int d\xi\psi(x-\xi,I-\eta,t)
K(\xi,\eta,t),
\end{equation}
with
\begin{equation}
 K(\xi,\eta,t)=\int^\infty_{-\infty}dj\int^\infty_0d\lambda
 e^{\frac{i}{\hbar}\left[\frac{1}{2\epsilon}\xi^2 -
     \epsilon\frac{m^2}{2}  +j\eta+(\overline{A_\mu\times I})
     \xi^\mu\right]},\label{eq:Kernel} 
\end{equation}
where
  \begin{eqnarray}
\overline{A_\mu\times I}\xi^\mu&=&\frac{1}{2}[A_\mu(x)\times
I+A_\mu(x-\xi)\times (I-\eta)]\xi^\mu \nonumber\\
&=&\left[A_\mu(x)\times I-\frac{1}{2} \xi^\nu\partial_\nu
  A_\mu(x)\times I -\frac{1}{2}A_\mu(x)\times\eta\right]\xi^\mu +
O(\xi^2\eta). 
\end{eqnarray}
%
Up to the second order in $\xi$ and the first order in $\eta$, the 
exponent in Eq.~(\ref{eq:Kernel}) can be expressed as
\begin{eqnarray}
&& \frac{i}{\hbar}\left[\frac{\xi^2}{2\epsilon} +
  j^a\left(A_\mu\times 
    I-\frac{1}{2}A_\mu\times\eta\right)^a\xi^\mu
-\frac{1}{2}j^a(\partial_\mu A_\nu\times
I)^a\xi^\mu\xi^\nu + j^a\eta^a - \epsilon\frac{m^2}{2}\right]
\nonumber \\
&=& \frac{i}{\hbar}\left(\frac{1}{2\epsilon}\bar{\xi}^\mu
G_{\mu\nu}\bar{\xi}^\nu-\frac{\epsilon}{2}a^\mu
G_{\mu\nu}a^\nu\right)
+ \frac{i}{\hbar}\left(j^a\eta^a - \epsilon\frac{m^2}{2}\right), 
\label{eq:Exponent in Kernel}
\end{eqnarray}
where we have defined
\begin{eqnarray}
&&G_{\mu\nu}\equiv \eta_{\mu\nu} -  \frac{1}{2} \epsilon j^a
[(\partial_\mu A_\nu\times I)^a + (\partial_\nu A_\mu\times I)^a] 
\nonumber,\\
&&a^\mu \equiv G^{\mu\nu}j^a[A_\nu\times
I-\frac{1}{2}A_\nu\times\eta]^a,\nonumber \\
&&\bar{\xi}^\mu\equiv\xi^\mu + \epsilon a^\mu,
\end{eqnarray}
and $G^{\mu\nu}$ is the matrix inverse of $G_{\mu\nu}\,,$ so that
$G_{\mu\nu}G^{\nu\lambda} = \delta_\mu^\lambda\,.$
On the right hand side of Eq.~(\ref{Eq:wavefunction}), we expand
the wave function $\psi(x-\xi,I-\eta,t)$ up to the second order in 
$\xi$ and $\eta$,
\begin{eqnarray}
\psi(x - \xi, I - \eta, t) &=& \psi(x, I, t)
-\xi^\mu \partial_\mu\psi - \eta^a\partial_a\psi
+ \xi^\mu\eta^a\partial_\mu\partial_a\psi \nonumber \\
&& + \frac{1}{2}\xi^\mu\xi^\nu\partial_\mu\partial_\nu\psi
+ \frac{1}{2}\eta^a\eta^b\partial_a\partial_b\psi.
\end{eqnarray}
Inclusion of higher orders $\xi$ and $\eta$ in a wave function
expansion brings higher order contribution in $\epsilon$, as we
will see below.  Now we are ready to integrate to get the wave
function at $t+\Delta t$,
\begin{eqnarray}
\psi(x,t+\Delta t)
=N\int d\lambda djd\eta
d\bar{\xi}&&\left[\psi- (\bar{\xi}^\mu - \epsilon
a^\mu) \partial_\mu\psi - \eta^a\partial_a\psi+
(\bar{\xi}^\mu-\epsilon a^\mu)\eta^a 
\partial_a\partial_\mu\psi\right.\nonumber\\
&&\left.+\frac{1}{2}\eta^a\eta^b \partial_a\partial_b\psi
+\frac{1}{2}(\bar{\xi}^\mu - \epsilon a^\mu)(\bar{\xi}^\nu 
- \epsilon a^\nu)\partial_\mu\partial_\nu\psi\right]\nonumber\\
&&\times e^{[\frac{i}{2\epsilon}\bar{\xi}^\mu
  G_{\mu\nu}\bar{\xi}^\nu- i\frac{\epsilon}{2} a^\mu
  G_{\mu\nu} a^\nu- i\epsilon\frac{m^2}{2} + ij^a\eta^a]/\hbar}.
\end{eqnarray}
Let us first perform the Gaussian integral for $\bar{\xi}$.  Odd
order terms in $\bar{\xi}$ vanish, so those are dropped and only
the terms up to first order in $\epsilon$ are kept in the result,
which reads
\begin{equation}
\psi(x,t+\Delta t)=N\int d\lambda dj d\eta d\bar{\xi} I_{\xi} 
e^{[- i\frac{\epsilon}{2} a^\mu
  G_{\mu\nu} a^\nu- i\epsilon\frac{m^2}{2} 
  + ij^a\eta^a]/\hbar}.
\label{eq:NI_xi}
\end{equation}
$I_{\xi}$ is defined as
\begin{eqnarray}
I_{\xi}&\equiv&\int
d\bar{\xi}\left[\psi +
\frac{1}{2}\bar{\xi}^\mu\bar{\xi}^\nu \partial_\mu\partial_\nu\psi
+ \epsilon a^\mu \partial_\mu\psi -
\eta^a\partial_a\psi+\frac{1}{2}\eta^a\eta^b \partial_a\partial_b\psi- 
\epsilon a^\mu\eta^a\partial_a\partial_\mu\psi\right]\nonumber\\ 
&&\qquad~~\times e^{\frac{i}{\hbar}\frac{1}{2\epsilon}\bar{\xi}^\mu
  G_{\mu\nu}\bar{\xi}^\nu} \nonumber \\
&=&N^{-1}(j)\left[\psi - \eta^a\partial_a\psi
  +\frac{1}{2}\eta^a\eta^b\partial_a\partial_b\psi\right.\nonumber\\  
&&\qquad\qquad\qquad\qquad\qquad~~\left.+
\epsilon\left(a^\mu\partial_\mu\psi
- a^\mu \eta^a \partial_a\partial_\mu\psi +
\frac{i\hbar}{2}G^{\mu\nu}\partial_\mu\partial_\nu\psi \right)
\right]\,.
\label{eq:Gaussian Integral}
\end{eqnarray}
In this equation we have written
\begin{eqnarray}
&&N^{-1}(j)=\int
d\bar{\xi}e^{\frac{i}{\hbar}\frac{1}{2\epsilon}\bar{\xi}^\mu
  G_{\mu\nu}\bar{\xi}^\nu} =
(2i\hbar\epsilon\pi)^{D/2}(\det{G_{\mu\nu}})^{-1/2},\nonumber \\
&&\int d\bar{\xi}\bar{\xi}^\mu\bar{\xi}^\nu
e^{\frac{i}{\hbar}\frac{1}{2\epsilon}\bar{\xi}^\mu
  G_{\mu\nu}\bar{\xi}^\nu}=i\hbar\epsilon N^{-1}(j)G^{\mu\nu}.
\end{eqnarray}
As mentioned earlier it can be seen that higher orders in $\xi$
bring higher orders in $\epsilon$. A similar statement is true
about $\eta$. Next for integration for $\eta$ and $j$ let us keep
terms only in the first order in $\epsilon$,
\begin{eqnarray}
e^{-\frac{i}{\hbar}\frac{\epsilon}{2} a^\mu G_{\mu\nu} a^\nu}
&=& 1-\frac{i\epsilon}{2\hbar}\left[(j\cdot(A_\mu\times I))^2
+ \frac{1}{4}(j\cdot(A_\mu\times \eta))^2-j\cdot(A_\mu\times I)
  j\cdot(A^\mu\times\eta)\right] \nonumber\\
&&+O(\epsilon^2),
\end{eqnarray}
where we have used the fact that $G_{\mu\nu}= \eta_{\mu\nu} +
O(\epsilon)$. Let us write the result as
\begin{equation}
\psi(x,t+\Delta t)=N \int\limits_0^\infty d\lambda I 
e^{-i\epsilon \frac{m^2}{2\hbar} },
\label{eq:psi NI}
\end{equation}
where
\begin{eqnarray}
I&=&\int djd\eta I_{\xi} e^{-\frac{i}{\hbar}\frac{\epsilon}{2}
a^\mu G_{\mu\nu} a^\nu + \frac{i}{\hbar}j\cdot\eta}\nonumber \\
&=&\int djd\eta N^{-1}(j)\left[\psi- \eta^a\partial_a\psi +
\frac{1}{2}\eta^a\eta^b\partial_a\partial_b\psi\right]
e^{\frac{i}{\hbar}j\cdot\eta}\nonumber \\
&&+\epsilon\int djd\eta N^{-1}(j)\left[- a^\mu\partial_\mu\psi
+ a^\mu\eta^a\partial_a\partial_\mu\psi\right]
e^{\frac{i}{\hbar}j\cdot\eta}  \nonumber\\
&&-\epsilon\frac{i}{2\hbar}\int djd\eta
N^{-1}(j)\left[(j\cdot(A_\mu\times I))^2 +
\frac{1}{4}(j\cdot(A_\mu\times \eta))^2 \right.\nonumber\\
&&\qquad\qquad\qquad\qquad\qquad\qquad\qquad\qquad
\left.-j\cdot(A_\mu\times I) 
j\cdot(A^\mu\times\eta)\right]\nonumber\\
&&\qquad\qquad\qquad\qquad\qquad\qquad\times\left[\psi -
  \eta^a\partial_a\psi + 
\frac{1}{2}\eta^a\eta^b\partial_a\partial_b\psi\right]
e^{\frac{i}{\hbar}j\cdot\eta}\nonumber \\
&&+\frac{i\hbar\epsilon}{2}\int djd\eta
N^{-1}(j)G^{\mu\nu}\partial_\mu\partial_\nu\psi
e^{\frac{i}{\hbar}j\cdot\eta} \nonumber \\
&=&I_1+I_2+I_3+I_4.
\end{eqnarray}
The integrals $I_1, \cdots, I_4$ are calculated in Appendix
\ref{Appx:Derivation of the Constraint Wave Equation}. Using the
results given there, we can write the final result
\begin{eqnarray}
I &=&N^{-1}(0)\psi+\lambda N^{-1}(0)\frac{i\hbar\Delta
  t}{2}\left[-(\partial_\mu
A^\mu\times I)^a\partial_a\psi-2(A^\mu\times
I)^a\partial_a\partial_\mu\psi\right.\nonumber \\
&&\left.+\frac{1}{4}\mbox{Tr}(A\cdot A)\psi+(A^\mu\times
I)^bf_{acb}A^c_\mu\partial_a\psi
+(A_\mu\times I)^a (A^\mu\times
I)^b\partial_a\partial_b\psi+\partial^2\psi\right],\;
\label{eq:I final}
\end{eqnarray}
where we have brought back $\lambda$ using $\epsilon=\lambda\Delta
t$, and written $\text{Tr}(A\cdot A)=f_{acb}A^c_\mu
f_{bda}A^{d\mu}$. If all the terms, including the mass term
$e^{-i\epsilon\frac{m^2}{2}}$, are now gathered, we find the wave
equation with respect to the parameter $t$. Let us first choose the
proportionality constant $N$ such that
\begin{eqnarray}
1=N\int^\infty_0d\lambda N^{-1}(0)&=&N(i2\hbar\Delta
t\pi)^{D/2}\int^\infty_0d\lambda
\lambda^{D/2}\nonumber\\
&=&\lim_{\Lambda\to\infty}
N\frac{(i2\hbar\Delta  t\pi)^{D/2}}{D/2+1}\Lambda^{D/2+1},
\end{eqnarray}
where $\Lambda$ is to be sent to infinity. That is,
\begin{equation}
N=\frac{D/2+1}{(i2\hbar\Delta
  t\pi)^{D/2}}\frac{1}{\Lambda^{D/2+1}}. 
\end{equation}
Going back to Eq.~(\ref{eq:I final}), we can write the wave
function $\psi(x,I,t+\Delta t)$ at $t+\Delta t$ as 
\begin{eqnarray}
&&\psi(x,I,t+\Delta t)-\psi(x,I,t)\nonumber\\
&=&\frac{i\hbar\Delta
  t}{2}N\int^\infty_0d\lambda\lambda N^{-1}(0)
\left[-(\partial_\mu A^\mu\times I)^a\partial_a\psi
-2(A^\mu\times I)^a\partial_a\partial_\mu\psi
+\frac{1}{4}\mbox{Tr}(A\cdot A)\psi\right.\nonumber\\
&&\left.\qquad\qquad+(A^\mu\times
I)^bf_{acb}A^c_\mu\partial_a\psi
+(A_\mu\times I)^a (A^\mu\times
I)^b\partial_a\partial_b\psi+\partial^2\psi
-\frac{m^2}{\hbar^2}\psi\right]\nonumber\\
&=&\frac{i\hbar\Delta
  t}{2}N\int^\infty_0d\lambda\lambda N^{-1}(0) \left[-(\partial_\mu  
A^\mu\times I)^a\frac{i}{\hbar}\hat{P}^a_I\psi-2(A^\mu\times
I)^a\left(\frac{i}{\hbar}\right)^2 
\hat{P}^a_I\hat{P}_\mu\psi\right.\nonumber\\
&&\qquad\qquad\qquad\qquad\qquad~~+\frac{1}{4} \mbox{Tr}(A\cdot A)
\psi+(A^\mu\times 
I)^bf_{acb}A^c_\mu\frac{i}{\hbar}\hat{P}^a_I\psi\nonumber\\
&&\left.\qquad\qquad\qquad~~+\left(\frac{i}{\hbar}\right)^2
  (A_\mu\times I)^a (A^\mu\times I)^b\hat{P}^a_I\hat{P}^b_I\psi+
  \left(\frac{i}{\hbar}\right)^2\hat{P}^2\psi- 
\frac{m^2}{\hbar^2}\psi\right],
\end{eqnarray}
where we have used that
$\partial^2=-\partial^2_0 + \partial^2_i =
-\frac{1}{\hbar^2}(-\hat{P}^2_0)-\frac{1}{\hbar^2}\hat{P}^2_i
=-\frac{1}{\hbar^2}\hat{P}^2$ and 
$\hat{P}^a_I=\frac{\hbar}{i}\partial_{I^a}$.
The wave equation with respect to the worldline parameter $t$ is 
\begin{eqnarray}
i\hbar\frac{\partial\psi}{\partial t}
&=&-\frac{1}{2}N\int^\infty_0d\lambda\lambda
N^{-1}(0)\left[-i\hbar(\partial_\mu A^\mu\times I)^a\hat{P}^a_I
  +2(A^\mu\times I)^a\hat{P}^a_I\hat{P}_\mu\right.\nonumber\\ 
&&\qquad\qquad\qquad\qquad\qquad+\frac{\hbar^2}{4} \mbox{Tr}(A\cdot
A)+i\hbar (A^\mu\times 
I)^bf_{acb}A^c_\mu\hat{P}^a_I\nonumber\\
&&\left.\qquad\qquad\qquad\qquad\qquad-(A_\mu\times I)^a
  (A^\mu\times 
I)^b\hat{P}^a_I\hat{P}^b_I-\hat{P}^2-m^2\right]\psi\nonumber\\
&=&-\frac{1}{2}N\int^\infty_0d\lambda\lambda
N^{-1}(0)\left[-(\hat{P}^2+m^2)
-2(A^a_\mu \hat{K}^a)\hat{P}^\mu\right.\nonumber\\
&&\left.\qquad\qquad\qquad\qquad\qquad-\eta^{\mu\nu}(A^a_\mu A^b_\nu)
\hat{K}^a\hat{K}^b-i\hbar(\partial_\mu
A^\mu\times I)^a\hat{P}^a_I\right.\nonumber\\
&&\left.\qquad\qquad\qquad\qquad\qquad+\frac{\hbar^2}{4}
  \mbox{Tr}(A\cdot A)+i\hbar 
(A^\mu\times I)^bf_{acb}A^c_\mu\hat{P}^a_I \right]\psi\nonumber\\
&=&\frac{1}{2}N\int^\infty_0d\lambda\lambda N^{-1}(0)[D\psi
+\Delta\psi ],
\label{eq:Time evolution eq(Schrodingers eq)}
\end{eqnarray}
where
\begin{eqnarray}
\hat{K}^a&=&-f_{bac}\hat{P}^b_II^c,
\label{eq:K-hat}\\
D\psi&=&[(\hat{P}_\mu+gA^a_\mu
\hat{K}^a)^2+m^2]_{\hat{Q}\hat{Q}\cdots\hat{P}\hat{P}}\psi,
\label{eq:Dpsi}\\
\Delta \psi&=&\left[i\hbar(\partial_\mu
A^\mu\times I)^a\hat{P}^a_I-\frac{\hbar^2}{4} \mbox{Tr}(A\cdot
A)-i\hbar (A^\mu\times
I)^bf_{acb}A^c_\mu\hat{P}^a_I\right]\psi. \label{eq:Deltapsi} 
\end{eqnarray}
We will see below that the operator $D+\Delta$ is an inverse Weyl
transform (the Wigner transform) of the classical Hamiltonian
$H=\frac{h}{2}[(P_\mu+gA^a_\mu K^a)^2+m^2]$.  The notation
$\hat{Q}\hat{Q}\cdots\hat{P}\hat{P}$ means that all the momentum
operators are put on the right while the position operators are on
the left. It can be seen that the right hand side in
Eq.~(\ref{eq:Time evolution eq(Schrodingers eq)}) is divergent,
since
\begin{eqnarray}
N\int^\infty_0d\lambda\lambda N^{-1}(0)
&=&N(i2\hbar\Delta
t\pi)^{D/2}\int^\infty_0d\lambda\lambda^{D/2+1}\nonumber\\
&=&\frac{D/2+1}{(i2\hbar\Delta
  t\pi)^{D/2}}\frac{1}{\Lambda^{D/2+1}}\frac{(i2\hbar\Delta
  t\pi)^{D/2}}{D/2+2}\Lambda^{D/2+2}\nonumber\\
&=&\frac{D/2+1}{D/2+2}\Lambda.
\end{eqnarray}
Eq.~(\ref{eq:Time evolution eq(Schrodingers eq)}) can be written as 
\begin{equation}
i\hbar\frac{\partial}{\partial t}\psi(x^\mu,I^a,t)
=\frac{1}{2}\frac{D/2+1}{D/2+2}\Lambda
(D\psi+\Delta\psi).\label{eq:Schrodinger eq}
\end{equation}
Dividing the both sides by the divergent factor and taking the
limit $\Lambda\to\infty$, we find 
\begin{equation}
D\psi+\Delta\psi=\lim_{\Lambda\to\infty}i\hbar\frac{D/2+2}{D/2+1}
\frac{1}{\Lambda}\frac{\partial}{\partial  
  t}\psi(x^\mu,I^a,t)=0.
  \label{wave-equation}
\end{equation}
This is the wave equation for $\psi$. For a fundamental
representation, the wave equation is obtained by replacing, e.g.,
$P^a_If_{abc}A^b_\mu I^c$ by $-iP^i_IT^b_{ij}A^b_{\mu}I^j$. Note
that in this substitution one has to maintain the order of indices
in $f_{abc}$.  Then in the fundamental representation
Eqs.~(\ref{eq:K-hat})-(\ref{eq:Deltapsi}) are replaced by
\begin{eqnarray}
\hat{K}^a&=&iT^a_{ij}\hat{P}^i_II^j,\\
D\psi&=&[(\hat{P}_\mu+igA^a_\mu T^a_{ij}I^j\hat{P}^{i}_I)^2+m^2
]_{\hat{Q}\hat{Q}\cdots\hat{P}\hat{P}}\psi,\\  
\Delta \psi&=&\left[-\hbar g(\partial_\mu
A^{b\mu}T^b_{ij}I^j)\hat{P}^i_I+\frac{\hbar^2}{4}
(T^a_{ij}T^b_{ji}A^a_\mu A^{b\mu})+i\hbar g^2(A^{a\mu}T^a_{ij}I^j)
T^b_{ki}A^b_\mu \hat{P}^k_I\right]\psi.
\label{eq:Deltapsi-fundamental} 
\end{eqnarray}

\subsection{Operator ordering}\label{sec:Operator Ordering}
In this subsection we verify the wave equation of
Eq.~(\ref{wave-equation}) by using a general mathematical formula
which relates the quantum Hamiltonian operator to the Hamiltonian
function in the path integral. The map from the operator to a
function in the phase space is called the Weyl transformation. The
inverse map is called the Wigner transformation. The correspondence
between the two spaces is one-to-one.
The Wigner transformation from a function $a(p,q)$ to an operator
$A(\hat{P},\hat{Q})$ is given in a compact form
as~\cite{Mizrahi:1975pw, McCoy:1932}.

\begin{eqnarray}
A(\hat{P},\hat{Q})=[e^{\frac{\hbar}{2i}\frac{\partial}{\partial
    p}\cdot\frac{\partial}{\partial
    q}}a(p,q)]_{p\rightarrow\hat{P},q\rightarrow\hat{Q}
  ;~\hat{Q}\hat{Q}\cdots \hat{P}\hat{P}},\label{eq:Wigner
  transformation}
\end{eqnarray}
where $\hat{Q}\hat{Q}\cdots \hat{P}\hat{P}$ means that all the
momentum operators are placed on the right side while the position
operators on the left side.

The classical Hamiltonian $H$ used in the path integral was
$\frac{h}{2}[(P_\mu+gA^a_\mu K^a)^2+m^2]$. The Wigner
transformation of that function on the phase space should give us
the appropriate operator to be used in the Schr\" odinger
equation. We note that the terms in $H$ for which there is an
ordering ambiguity are $2P^\mu A^a_\mu K^a$ and
$\eta^{\mu\nu}A^a_\mu K^aA^b_\nu K^b$. Plugging these functions
into Eq.~(\ref{eq:Wigner transformation}), we get
\begin{equation}
e^{\frac{\hbar}{2i}\frac{\partial}{\partial
    P_\mu}\frac{\partial}{\partial x^\mu}}(2P^\mu A^a_\mu K^a)
\Rightarrow2A\hat{K}\hat{P}+\frac{\hbar}{i}\partial_\mu A^\mu
\hat{K}
\label{eq:Wigner-PAK}
\end{equation}
and
\begin{eqnarray}
&&e^{\frac{\hbar}{2i}\frac{\partial}{\partial
    P^a_I}\frac{\partial}{\partial I^a}}(\eta^{\mu\nu}A^a_\mu
K^aA^b_\nu K^b)\\
&\Rightarrow& (A_\mu\times I)^a(A^\mu\times
I)^b\hat{P}^a_I\hat{P}^b_I\nonumber\\ 
&&+A^a\cdot A^b\frac{\hbar}{2i}\frac{\partial}{\partial
  P^g_I}\frac{\partial}{\partial
  I^g}(f_{acd}f_{bef}P^c_II^dP^e_II^f)\nonumber\\ 
&&-\frac{1}{2}A^a\cdot A^b\frac{\hbar^2}{4}
\frac{\partial}{\partial P^g_I}\frac{\partial}{\partial
  I^g}\frac{\partial}{\partial 
  P^h_I}\frac{\partial}{\partial
  I^h}(f_{acd}f_{bef}P^c_II^dP^e_II^f)\nonumber\\ 
&=&(A_\mu\times I)^a(A^\mu\times
I)^b\hat{P}^a_I\hat{P}^b_I\nonumber\\ 
&&+\frac{\hbar}{2i}(-2A^c_\mu f_{bca}(A^\mu\times
I)^b)\hat{P}^a_I-\frac{1}{2}\frac{\hbar^2}{4}2(f_{acb}A^c)
\cdot(f_{bda}A^d))\nonumber\\   
&=&(A_\mu\times I)^a(A^\mu\times
I)^b\hat{P}^a_I\hat{P}^b_I-\frac{\hbar}{i}A^c_\mu
f_{bca}(A^\mu\times
I)^b\hat{P}^a_I-\frac{\hbar^2}{4}\mbox{Tr}(A\cdot A).
\label{eq:Wigner-AKAK} 
\end{eqnarray}
The additional pieces, $\frac{\hbar}{2i}\partial_\mu A^\mu
\hat{K}$, $-\frac{\hbar}{i}A^c_\mu f_{bca}(A^\mu\times
I)^b\hat{P}^a_I$ and $-\frac{\hbar^2}{4}\mbox{Tr}[A\cdot A]$,
exactly match with the terms in Eq.~(\ref{eq:Deltapsi}), the
expression for $\Delta\psi$.
Let us rewrite the expressions (\ref{eq:Wigner-PAK}) and
(\ref{eq:Wigner-AKAK}) in a different form.
The operator (\ref{eq:Wigner-PAK}) can be written as
\begin{equation}
e^{\frac{\hbar}{2i}\frac{\partial}{\partial
    P_\mu}\frac{\partial}{\partial x^\mu}}2PAK
\Rightarrow(\hat{P}\cdot A^a+A^a\cdot\hat{P})K^a\,.
\end{equation}
Also, using the relations
\begin{eqnarray}
[(A_\mu\times
I)^b,\hat{P}^a_I]&=&gf_{bcd}A^c_\mu[I^d,\hat{P}^a_I]\nonumber\\ 
&=&gf_{bcd}A^c_\mu\delta^{da}=i\hbar gf_{bca}A^c_\mu, 
\end{eqnarray}
and
\begin{eqnarray}
(A_\mu\times I)^a(A^\mu\times I)^b\hat{P}^a_I\hat{P}^b_I
&=&(A_\mu\times I)^a[(A_\mu\times I)^b,\hat{P}^a_I]\hat{P}^b_I
+(A_\mu\times I)^a\hat{P}^a_I(A_\mu\times I)^b\hat{P}^b_I\nonumber\\
&=&(A_\mu\times I)^a(i\hbar f_{bca}A^c_\mu)\hat{P}^b_I
+(A_\mu\times I)^a\hat{P}^a_I(A_\mu\times I)^b\hat{P}^b_I\nonumber\\
&=&(A_\mu\times I)^a(i\hbar f_{bca}A^c_\mu)\hat{P}^b_I
+(A^a_\mu\hat{K}^aA^{b\mu}\hat{K}^b)\nonumber\\
&=&(A_\mu\times I)^b(i\hbar f_{acb}A^c_\mu)\hat{P}^a_I
+(A^a_\mu\hat{K}^aA^{b\mu}\hat{K}^b)\nonumber\\
&=&-(A_\mu\times I)^b(i\hbar f_{bca}A^c_\mu)\hat{P}^a_I
+(A^a_\mu\hat{K}^aA^{b\mu}\hat{K}^b),
\end{eqnarray}
we can rewrite Eq.~(\ref{eq:Wigner-AKAK}) as
\begin{eqnarray}
e^{\frac{\hbar}{2i}\frac{\partial}{\partial
    P^a_I}\frac{\partial}{\partial I^a}}(\eta^{\mu\nu}A^a_\mu
K^aA^b_\nu K^b)&\Rightarrow&-(A_\mu\times I)^b(i\hbar
f_{bca}A^c_\mu)\hat{P}^a_I
+(A^a_\mu\hat{K}^a)(A^{b\mu}\hat{K}^b)\nonumber\\
&&+i\hbar A^c_\mu f_{bca}(A^\mu\times I)^b\hat{P}^a_I
-\frac{\hbar^2}{4}\mbox{Tr}(A\cdot A)\nonumber\\
&=&(A^a_\mu\hat{K}^aA^{b\mu}\hat{K}^b) -
\frac{\hbar^2}{4}\mbox{Tr}(A\cdot A).
\end{eqnarray}
Thus the Hamiltonian operator can be written without assigning 
a particular ordering,
\begin{equation}
\hat{H}=\frac{h}{2}\left[(\hat{P}_\mu+gA^a_\mu\hat{K}^a)^2+m^2 -
  g^2\frac{\hbar^2}{4}\mbox{Tr}(A\cdot A)\right].
\label{unordered-H}
\end{equation}
We clearly see that the Hamiltonian operator consists of the
operator naively replaced by the classical Hamiltonian and the
gauge non-invariant term $g^2\frac{\hbar^2}{4}\mbox{Tr}(A\cdot A)$.
As far as the operators orderings are relevant, there is no reason
to expect that the Hamiltonian function in the path integral should
correspond to operator in which the phase space variables have been
naively substituted by the corresponding quantum operators.  We
note that the last term has been previously found in the
literature~\cite{Fresneda:2007vc}.

The Hamiltonian commutes with the internal angular momentum $K^2$,
so one can factor out the eigenfunction for $K^2$. Because of using
real and bosonic variables, $K^a$ behaves as an ordinary angular
momentum and hence has integer eigenvalues.

\section{Conclusion}

In this paper we have considered the question: What is a classical
non-Abelian point particle? More specifically, what is the
classical dynamics of such a particle in a background non-Abelian
gauge field? We started from a classical action describing the
position of the particle as well as its charge, described by a
dynamical `internal' vector in some representation of the gauge
group. We found that when this internal vector is in either the
adjoint or the fundamental representation, the charge vector that
enters the generalized Lorentz force equation is an adjoint vector
constructed from the original charge vector and its conjugate
momentum. So the charge vector that determines the space-time
trajectory of the particle is in the adjoint representation in both
cases.

The equations were originally derived from quantum theory, so we
decided to quantize the classical action in the path integral
formalism as a kind of cross check. Using the worldline formalism
so as to include particles of zero mass, we found that the sum over
paths includes only those paths along which the internal charge
vector is parallel transported. We also derived a wave equation
from this path integral and showed that the Hamiltonian operator in
the wave equation exactly matches with the Hamiltonian operator
transformed from the classical Hamiltonian function by the Weyl
correspondence.

There are however some differences between the quantum theory of a
non-Abelian charged field and that of a non-Abelian point particle
as constructed from the classical action. That is not a failure of
quantization, nor a contradiction with established knowledge. The
first point of departure is the fact that the classical `isospin',
the charge vector in the action, is a continuous variable, whereas
for quantum particles, the isospin $\vec I$ is quantized, with
fixed ${\vec I}^2$. This can be resolved in the following way. The
charge vector ${\vec K}$ in our setup corresponds to the isospin
vector operator upon quantization. By construction the vector
${\vec K}$ is like an angular momentum operator in the internal
vector space, so it will have discrete eigenvalues when the theory
is quantized.  The quantum Hamiltonian commutes with ${\vec K}^2$
as can be easily checked, so a particle in a given isospin
eigenstate remains in that eigenstate. Then the quantum particle
with a fixed isospin corresponds to the particle in an eigenstate
of ${\vec K}^2$. However, since ${\vec K}$ corresponds to $\vec x
\times \vec p$ in the internal space, it can have only integer
eigenvalues. This is analogous to there being no truly classical
description of half-integer spin. The solution would be to use
anti-commuting variables for the internal charge vector. That would
allow, for example, half-integer isospins when the gauge group is
SU(2).

Another discrepancy is the appearance of the term $\frac14 g^2
\hbar^2\text{Tr}(A\cdot A)$ in the quantum Hamiltonian operator,
Eq.~(\ref{unordered-H}). This term breaks the gauge symmetry
present in the classical Hamiltonian, although the Hamiltonian is
still symmetric under {\em constant} internal rotations. This term
appears to be a genuine effect of quantization: the Hamiltonian
operator as derived from the path integral exactly matches with the
Hamiltonian operator constructed from the classical Hamiltonian
function by the Weyl correspondence. While a Hamiltonian
constructed directly from a non-Abelian gauge symmetric quantum
field theory is not expected to contain such a term, we note that
the anomalous term is not unknown in the
literature~\cite{Fresneda:2007vc}.

The source of this discrepancy is the following. The constraint
which implements gauge transformations comes from the Hamiltonian
for the gauge field, but we have treated the non-Abelian gauge
field $A_\mu$ as a background field, ignoring its dynamics. We
could of course try to include the Lagrangian for the gauge
field. But it is known that for an electrically charged point
particle in a background electromagnetic field the joint action
leads to inconsistencies, stemming from the fact that the field due
to the charged particle itself diverges at the position of the
latter~\cite{Rohrlich:1965bk}. The problem is resolved by starting
from the Lorentz-Dirac equation instead of the Lorentz force
equation, and making further modifications so that quantization
leads to the Dirac equation~\cite{Barut:1984prl, Kar:2010jm}.

The appearance of the gauge symmetry breaking term in the quantum
Hamiltonian for the non-Abelian point particle is related to this
classical difficulty of defining the action of a point particle in
a dynamical gauge field. In the Hamiltonian picture, the constraint
which implements gauge symmetry appears only if the gauge fields
have their own dynamics. But if the non-Abelian point particle is
coupled to a dynamical gauge field, the radiation reaction must be
included as for the ordinary electric charge. In that case, Wong's
generalization of the Lorentz force equation will have to be
replaced by something analogous to the Lorentz-Dirac equation, and
a corresponding action, as the starting point.

\begin{acknowledgments}
This work was supported in part by the Dept. of Science and
Technology, Govt. of India through a grant
No. SR/S2/HEP-0006/2008. 
\end{acknowledgments}

\appendix
\appendixpage
\addappheadtotoc

\section{Derivation of the equation of
  motion}\label{Appx:Derivation of the Equation of Motion} 

In this appendix we show that by varying the path integral in
Eq.~(\ref{eq:final path integral}) with respect to $x^\mu$, we can
recover the classical equation of motion for $x^\mu$ from the
vanishing first order variation,
\begin{equation}
0=\delta\int^\infty_{-\infty}dj\int^\infty_0d\lambda\mathscr
{D}h\delta(h-\lambda)\int\mathscr{D}x^\mu e^{\frac{i}{\hbar}\int
  dt[\frac{1}{2h}\dot{x}^2-h\frac{m^2}{2}] 
+\frac{i}{\hbar}j(I_fTe^{-\int^{t_f}_{t_i}  Adt}-I_i)},
\label{eq:varying final path integral} 
\end{equation}
where $A_{ac}\equiv gf_{abc}A^b_\mu\dot{x}^\mu$.
Using the relation
\begin{equation}
\delta U(t_b,t_a)=-\int^{t_b}_{t_a}dt^\prime U(t_b,t^\prime)\delta
A(t^\prime)U(t^\prime,t_a), 
\end{equation}
where $U(t_b,t_a)=Te^{-\int^{t_b}_{t_a} Adt}$, we can write
\begin{eqnarray}
\frac{\delta}{\delta x^\mu(t)}
U(t_f,t_i)&=&-\int^{t_f}_{t_i}dt^\prime
U(t_f,t^\prime)\frac{\delta}{\delta x^\mu(t)}
A(t^\prime)U(t^\prime,t_i)\nonumber\\ 
&=&-\int^{t_f}_{t_i}dt^\prime
U(t_f,t^\prime)[\delta(t^\prime-t)\partial_\mu
A_\nu(t^\prime)\dot{x}^\nu (t^\prime) 
-A_\mu\frac{d}{dt}\delta(t^\prime-t)]U(t^\prime,t_i)\nonumber\\
&=&-U(t_f,t)\partial_\mu A_\nu(t)\dot{x}^\nu U(t,t_a)
-\frac{d}{dt}[-U(t_f,t)A_\mu U(t,t_i)]\nonumber\\
&=&-U(t_f,t)[\partial_\mu A_\nu(t)\dot{x}^\nu]U(t,t_i)\nonumber\\ 
&&+U(t_f,t)A_\nu \dot{x}^\nu A_\mu U(t,t_i)+U(t_f,t)\partial_\nu
A_\mu\dot{x}^\nu U(t,t_i)\nonumber\\ 
&&-U(t_f,t)A_\mu A_\nu \dot{x}^\nu U(t,t_i)\nonumber\\
&=&-U(t_f,t)F_{\mu\nu}\dot{x}^\nu U(t,t_i),
\end{eqnarray}
where $[F_{\mu\nu}]_{ac}\equiv gf_{abc}[\partial_\mu
A^b_\nu-\partial_\nu A^b_\mu+gf_{bde}A^d_\mu A^e_\nu]$. Recalling
that it has been defined in Eq.~(\ref{eq:final path integral}) that
$j^a=P^a_{I}(t_i)$, the relevant term $jI_f\frac{\delta}{\delta
  x^\mu(t)} U(t_f,t_i)$ can be written 
\begin{eqnarray}
P^a_{I}(t_i)[-I_fU(t_f,t)F_{\mu\nu}\dot{x}^\nu U(t,t_i)]_a
&=&P^a_{I}(t_i)[-I(t)F_{\mu\nu}\dot{x}^\nu U(t,t_i)]_a\nonumber\\
&=&[-I(t)F_{\mu\nu}\dot{x}^\nu U(t,t_i)]_aP^a_{I}(t_i)\nonumber\\
&=&-I(t)F_{\mu\nu}\dot{x}^\nu P_{I}(t)\nonumber\\
&=&-I^a(t)gf_{abc}F^b_{\mu\nu}\dot{x}^\nu P^c_{I}(t)\nonumber\\
&=&-gK^bF^b_{\mu\nu}\dot{x}^\nu,
\end{eqnarray}
where $K^a=f_{abc}P^b_{I}I^c$. Therefore, the vanishing first order
in $\delta x^\mu$ in Eq.~(\ref{eq:varying final path integral})
leads to the equation of motion for $x^\mu$. 
\begin{eqnarray}
&&\int
dt[-\frac{d}{dt}\left(h^{-1}\dot{x}_\mu\right) +
J^a_i[-I_fU(t_f,t)F_{\mu\nu}\dot{x}^\nu  
U(t,t_i)]_a\delta x^\mu\nonumber\\ 
&=&\int dt[-\frac{d}{dt}\left(\frac{m\dot{x}_\mu}
  {\sqrt{-\dot{x}^2}}\right)-gK^bF^b_{\mu\nu}\dot{x}^\nu]\delta 
x^\mu\nonumber\\ 
&=&\int dt[-\frac{d}{dt}\left(\frac{m\dot{x}_\mu}
  {\sqrt{-\dot{x}^2}}\right)-gK^bF^b_{\mu\nu}\dot{x}^\nu]\delta
x^\mu\nonumber\\ 
&\Rightarrow&\frac{d}{dt}\left(\frac{m\dot{x}_\mu}
  {\sqrt{-\dot{x}^2}}\right)+gK^bF^b_{\mu\nu}\dot{x}^\nu=0.\\ 
\end{eqnarray}

\section{Derivation of the constrained wave
  equation}\label{Appx:Derivation of the Constraint Wave Equation} 

The kernel was written from Eq.~(\ref{eq:Exponent in Kernel}) in the
following notation,
\begin{equation}
K(\xi,\eta,t)
=\int^\infty_{-\infty}dj\int^\infty_0d
\lambda\exp{\left\{\frac{i}{\hbar}  
\left[\frac{1}{2\epsilon}\bar{\xi}^\mu
  G_{\mu\nu}\bar{\xi}^\nu-\frac{\epsilon}{2}a^\mu G_{\mu\nu}a^\nu 
+j\eta-\epsilon\frac{m^2}{2}\right]\right\}},
\end{equation}
where
\begin{eqnarray}
&&G_{\mu\nu}\equiv \eta_{\mu\nu} -  \frac{1}{2} \epsilon j^a
[(\partial_\mu A_\nu\times I)^a + (\partial_\nu A_\mu\times I)^a]
\nonumber,\\
&&a^\mu \equiv G^{\mu\nu}j^a[A_\nu\times
I-\frac{1}{2}A_\nu\times\eta]^a,\nonumber \\
&&\bar{\xi}^\mu\equiv\xi^\mu + \epsilon a^\mu,
\end{eqnarray}
The wave function $\psi(x,t+\Delta t)$ at $t+\Delta t$ is a sum 
of the wave functions $\psi(x-\xi,I-\eta,t)$ with the weight
$K(\xi,\eta,t)$. 
\begin{eqnarray}
\psi(x,t+\Delta t)
&=&N\int d\eta\int d\xi\psi(x-\xi,I-\eta,t)
K(\xi,\eta,t)\nonumber\\ 
&=&N\int d\lambda djd\eta d\xi\left(\psi-\partial_\mu\psi\xi^\mu 
-\partial_a\psi\eta^a
+\frac{1}{2}\partial_\mu\partial_\nu\psi\xi^\mu\xi^\nu 
\right.\nonumber\\
&&\left.\qquad\qquad\qquad\qquad\qquad\qquad\qquad~~+
  \frac{1}{2}\partial_a\partial_b\psi\eta^a\eta^b  
+\partial_a\partial_\mu\psi\xi^\mu\eta^a\right)\nonumber\\
&&\qquad\qquad\qquad\qquad\qquad\qquad\times 
e^{\frac{i}{2\hbar\epsilon}\bar{\xi}^\mu 
G_{\mu\nu}\bar{\xi}^\nu}e^{-\frac{i\epsilon}{2\hbar}a^\mu
G_{\mu\nu}a^\nu-\frac{i\epsilon}{2\hbar}m^2
+\frac{i}{\hbar}j\eta}\nonumber\\  
&=&N\int d\lambda djd\eta
d\bar{\xi}\left[\psi-\partial_\mu\psi(\bar{\xi}^\mu -\epsilon
  a^\mu)-\partial_a\psi\eta^a
  +\partial_a\partial_\mu\psi(\bar{\xi}^\mu- \epsilon
  a^\mu)\eta^a\right.\nonumber\\ 
&&\left.\qquad\qquad\qquad~+
  \frac{1}{2}\partial_a\partial_b\psi\eta^a\eta^b+
  \frac{1}{2}\partial_\mu\partial_\nu\psi(\bar{\xi}^\mu-\epsilon   
  a^\mu)(\bar{\xi}^\nu-\epsilon a^\nu) 
\right]\nonumber\\
&&\qquad\qquad\qquad\qquad\qquad~~\times
e^{\frac{i}{2\hbar\epsilon}\bar{\xi}^\mu
  G_{\mu\nu}\bar{\xi}^\nu}e^{-\frac{i\epsilon}{2\hbar}a^\mu
  G_{\mu\nu}a^\nu-\frac{i\epsilon}
  {2\hbar}m^2+\frac{i}{\hbar}j\eta}\nonumber\\  
&=& N\int d\lambda dj d\eta d\bar{\xi} I_{\xi} 
e^{[- i\frac{\epsilon}{2} a^\mu
  G_{\mu\nu} a^\nu- i\epsilon\frac{m^2}{2} 
  + ij^a\eta^a]/\hbar},
\end{eqnarray}
which was Eq.~(\ref{eq:NI_xi}). Let us perform the Gaussian
integral $I_{\xi}$ first. In the end we are only interested in the
first orders in $\epsilon$ after suitable choice of normalization
factor $N$.
\begin{eqnarray}
I_{\xi}&=&\int
d\bar{\xi}\left[\psi-\partial_\mu\psi(\bar{\xi}^\mu-\epsilon
  a^\mu)-\partial_a\psi\eta^a 
+\frac{1}{2}\partial_\mu\partial_\nu\psi(\bar{\xi}^\mu-\epsilon
a^\mu)(\bar{\xi}^\nu-\epsilon a^\nu)\right.\nonumber\\ 
&&\left.\qquad\qquad\qquad\qquad\qquad~~
  +\frac{1}{2}\partial_a\partial_b\psi\eta^a\eta^b  
-\partial_a\partial_\mu\psi(\bar{\xi}^\mu-\epsilon a^\mu)
\eta^a\right]e^{\frac{i}{2\hbar\epsilon}\bar{\xi}^\mu
G_{\mu\nu}\bar{\xi}^\nu}\nonumber\\ 
&=&\int d\bar{\xi}\left[\psi-\partial_a\psi\eta^a +
  \frac{1}{2}\partial_a\partial_b\psi\eta^a\eta^b  
+\epsilon\partial_\mu\psi(a^\mu)-\epsilon\partial_a\partial_\mu\psi(
a^\mu)\eta^a\right]e^{\frac{i}{2\hbar\epsilon}\bar{\xi}^\mu
G_{\mu\nu}\bar{\xi}^\nu}\nonumber\\ 
&&+\int
d\bar{\xi}\left(\frac{1}{2}\partial_\mu\partial_\nu
  \psi\bar{\xi}^\mu\bar{\xi}^\nu\right)
e^{\frac{i}{2\hbar\epsilon}\bar{\xi}^\mu  
  G_{\mu\nu}\bar{\xi}^\nu}\nonumber\\ 
&=&N^{-1}(j)\left(\psi-\partial_a\psi\eta^a
+\frac{1}{2}\partial_a\partial_b\psi\eta^a\eta^b\right)
+N^{-1}(j)\epsilon[\partial_\mu\psi(a^\mu)-\partial_a\partial_\mu\psi(
a^\mu)\eta^a]\nonumber\\ 
&&+N^{-1}(j)\frac{i\hbar\epsilon}{2}G^{\mu\nu} 
\partial_\mu\partial_\nu\psi,\label{eq:I xi} 
\end{eqnarray}
where it has been defined
\begin{equation}
N^{-1}(j)=\int d\bar{\xi}e^{\frac{i}{2\hbar\epsilon} \bar{\xi}^\mu 
  G_{\mu\nu}\bar{\xi}^\nu}=
(i2\hbar\epsilon\pi)^{D/2}(\det{G_{\mu\nu}})^{-1/2}.   
\end{equation}
The vanishing exponential integrals with odd multiples in
$\bar{\xi}$ have been dropped out and using the formulae $\delta
(\det{G_{\mu\nu}})=(\det{G_{\mu\nu}})G^{\mu\nu}\delta G_{\mu\nu}$
and $\delta G^{\mu\nu}=-G^{\mu\rho}G^{\nu\sigma}\delta
G_{\rho\sigma}$, the following integral has been evaluated.
\begin{eqnarray}
\int d\bar{\xi}\bar{\xi}^\mu\bar{\xi}^\nu
e^{\frac{i}{2\hbar\epsilon}\bar{\xi}^\mu G_{\mu\nu}\bar{\xi}^\nu} 
&=&-i2\hbar\epsilon\frac{\partial N^{-1}(j)} {\partial
  G_{\mu\nu}}\nonumber\\ 
&=&-i2\hbar\epsilon(i2\hbar\epsilon\pi)^{D/2}\frac{\partial
  G^{-1/2}}{\partial G_{\mu\nu}}\nonumber\\ 
&=&-i2\hbar\epsilon(i2\hbar\epsilon\pi)^{D/2}
\left[-\frac{1}{2}(\det{G_{\mu\nu}})^{-3/2}\right]
(\det{G_{\mu\nu}})G^{\mu\nu}\nonumber\\  
&=&i\hbar\epsilon N^{-1}(j)G^{\mu\nu}.
\end{eqnarray}
Next what remains is to integrate over $(\eta,j)$.
\begin{equation}
I=\int djd\eta I_{\xi}e^{-\frac{i\epsilon}{2\hbar}a^\mu
  G_{\mu\nu}a^\nu+\frac{i}{\hbar}j\eta}. 
\end{equation}
Since only the first order in $\epsilon$ is needed, in the
exponential $e^{-\frac{i\epsilon}{2\hbar}a^\mu G_{\mu\nu}a^\nu}$
only the terms up to first order in $\epsilon$ are kept.
\begin{eqnarray}
&&e^{-\frac{i\epsilon}{2\hbar}a^\mu
  G_{\mu\nu}a^\nu+\frac{i}{\hbar}j\eta} \nonumber\\
&=&\left[1-\frac{i\epsilon}{2\hbar} 
j\left(A_\mu\times I-\frac{1}{2}A_\mu\times\eta\right)G^{\mu\nu}j
\left(A_\nu\times
  I-\frac{1}{2}A_\nu\times\eta+\cdots\right)\right]
e^{\frac{i}{\hbar}j\eta}\nonumber\\  
&=&\left[1-\frac{i\epsilon}{2\hbar}
j\left(A_\mu\times
  I-\frac{1}{2}A_\mu\times\eta\right)\eta^{\mu\nu}j 
\left(A_\nu\times
  I-\frac{1}{2}A_\nu\times\eta\right)+O(\epsilon^2)\right]
e^{\frac{i}{\hbar}j\eta}\nonumber\\  
&=&\left[1-\frac{i\epsilon}{2\hbar}\left(j\left(A_\mu\times I\right)
 j\left(A^\mu\times I\right)
+\frac{1}{4}j\left(A_\mu\times \eta\right) j\left(A^\mu\times
  \eta\right) 
-j\left(A_\mu\times I\right) j\left(A^\mu\times
  \eta\right)\right)\right] 
e^{\frac{i}{\hbar}j\eta}\nonumber\\
&&+O(\epsilon^2),
\end{eqnarray}
where $G_{\mu\nu}=\eta_{\mu\nu}+O(\epsilon)$ has been used. The
function $e^{\frac{i}{\hbar}j\eta}$ is proven to be useful since a
function of $\eta$ can be converted to a differential operator with
respect to $j$.
\begin{eqnarray}
\int dj d\eta e^{\frac{i}{\hbar}j\eta}f(\eta)g(j)
&=&\int dj d\eta g(j)f\left(\frac{\hbar}{i}\frac{\partial}{\partial
    j}\right)e^{\frac{i}{\hbar}j\eta}\nonumber\\ 
&=&\int dj d\eta
e^{ij\eta}f\left(-\frac{\hbar}{i}\frac{\partial}{\partial
    j}\right)g(j)\nonumber\\ 
&=&(2\pi\hbar)^n\int dj
\delta(j)f\left(-\frac{\hbar}{i}\frac{\partial}{\partial
    j}\right)g(j)\nonumber\\ 
&=& (2\pi\hbar)^nf\left(-\frac{\hbar}{i}\frac{\partial}{\partial
    j}\right)g(j)|_{j=0}, 
\end{eqnarray}
where $n$ in $(2\pi\hbar)^n$ is dimension of $j^a$. Note that we
will omit $(2\pi\hbar)^n$ in later calculations since it can be
absorbed into the normalization factor $N$. It is worthwhile to
recognize two properties for calculational convenience. The first
is that to get non-vanishing terms the number of powers in $\eta$
must be the same as that of $j$s. If $k\neq m$, 
\begin{equation}
 \int dj d\eta e^{\frac{i}{\hbar}j\eta}\eta^kj^m=0
\end{equation}
The second is that $n$th derivative of $N^{-1}(j)$ with respect $j$
brings $\epsilon^n$ order. For instance, 
\begin{eqnarray}
\frac{\partial N^{-1}(j)}{\partial j^a}
&=&(i2\epsilon\hbar\pi
)^{D/2}\left[-\frac{1}{2}(\det{G_{\mu\nu}})^{-3/2}\right]
(\det{G_{\mu\nu}})G^{\mu\nu}\frac{\partial 
  G_{\mu\nu}}{\partial j^a}\nonumber\\ 
&=&-\frac{1}{2}N^{-1}(j)G^{\mu\nu}\frac{\partial
  G_{\mu\nu}}{\partial j^a} =\frac{\epsilon}{2}
N^{-1}(j)G^{\mu\nu}B^a_{\mu\nu}. \label{eq:Derivative of N inverse} 
\end{eqnarray}
It can be easily seen that the further derivatives also have the
similar properties. This implies that higher than first derivatives
of $N^{-1}(j)$ with respect to $j$ can be discarded because we want
to keep terms only up to the first order in $\epsilon$. The final
integral $I$ is written in four pieces. 
\begin{equation}
I=\int djd\eta I_{\xi}e^{-\frac{i\epsilon}{2\hbar}a^\mu
  G_{\mu\nu}a^\nu+\frac{i}{\hbar}j\eta} 
=I_1+I_2+I_3+I_4,
\end{equation}
where
\begin{eqnarray}
I_1&=&\int djd\eta N^{-1}(j)\left(\psi-\partial_a\psi\eta^a
+\frac{1}{2}\partial_a\partial_b\psi\eta^a\eta^b\right)
e^{\frac{i}{\hbar}j\eta},\\  
I_2&=&\epsilon\int djd\eta N^{-1}(j)[\partial_\mu\psi(a^\mu)
-\partial_a\partial_\mu\psi(
a^\mu)\eta^a]e^{\frac{i}{\hbar}j\eta},\\ 
I_3&=&-\frac{i\epsilon}{2\hbar}\int djd\eta N^{-1}(j)
\left(j\left(A_\mu\times I\right) j\left(A^\mu\times I \right) 
+\frac{1}{4}j\left(A_\mu\times \eta\right) j\left(A^\mu\times
  \eta\right)\right.\nonumber\\ 
&&\left.\qquad\qquad\qquad\qquad\qquad\qquad\qquad\qquad\qquad
-j\left(A_\mu\times I\right) j\left(A^\mu\times
  \eta\right)\right)\nonumber\\ 
&&\qquad\qquad\qquad\qquad~~\times\left(\psi- \partial_a\psi\eta^a
  + \frac{1}{2}\partial_a\partial_b\psi\eta^a\eta^b\right)  
e^{\frac{i}{\hbar}j\eta},\\
I_4&=&\frac{i\hbar\epsilon}{2}\int djd\eta
N^{-1}(j)G^{\mu\nu}\partial_\mu\partial_\nu\psi
e^{\frac{i}{\hbar}j\eta}. 
\end{eqnarray}
Let us evaluate the first integral $I_1$.
\begin{eqnarray}
I_1&=&\int djd\eta
N^{-1}(j)\left(\psi-\partial_a\psi\eta^a +
  \frac{1}{2}\partial_a\partial_b\psi\eta^a\eta^b\right)  
e^{\frac{i}{\hbar}j\eta}\nonumber\\
&=&\int djd\eta
N^{-1}(j)\left(\psi-\partial_a\psi\frac{\hbar}{i}
  \frac{\partial}{\partial j^a}\right)e^{\frac{i}{\hbar}j\eta}, 
\end{eqnarray}
where $\frac{1}{2}\partial_a\partial_b\psi\eta^a\eta^b$ has been
dropped out since it corresponds to the second derivative of
$N^{-1}(j)$ with respect to $j$, which is in the second order in
$\epsilon$. Thus, using Eq.~(\ref{eq:Derivative of N inverse}) 
\begin{eqnarray}
I_1&=&N^{-1}(0)\psi+\partial_a\psi\frac{\hbar}{i}
\frac{\partial}{\partial j^a}N^{-1}(j)|_{j=0}\nonumber\\  
&=&N^{-1}(0)\psi+\partial_a\psi \frac{\hbar\epsilon}{2i}
N^{-1}(0)G^{\mu\nu}B^a_{\mu\nu}\nonumber\\ 
&=&N^{-1}(0)\left(\psi- \frac{i\hbar\epsilon}{2}
\partial_a\psi\eta^{\mu\nu}B^a_{\mu\nu}\right)\nonumber\\    
&=&N^{-1}(0)\left[\psi-\frac{i\hbar\epsilon}{2} (\partial_\mu
  A^\mu\times I)^a\partial_a\psi\right]. 
\end{eqnarray}
Next, in $I_2$ we discard $j\eta^{\mu\nu}(A_\nu\times I)$ and
$\frac{1}{2}j\eta^{\mu\nu}(A_\nu\times\eta)\eta^a$ since they do
not have the same powers in $j$ and $\eta$, so those terms vanish
or contribute to higher order in $\epsilon$. 
\begin{eqnarray}
I_2&=&\epsilon\int djd\eta
N^{-1}(j)[\partial_\mu\psi(a^\mu)-\partial_a\partial_\mu\psi(
a^\mu)\eta^a]e^{\frac{i}{\hbar}j\eta}\nonumber\\ 
&=&\epsilon\int djd\eta N^{-1}(j)\left[\partial_\mu\psi
  jG^{\mu\nu}\left(A_\nu\times
    I-\frac{1}{2}A_\nu\times\eta\right)\right.\nonumber\\ 
&&\left.\qquad\qquad\qquad\qquad\qquad-\partial_a\partial_\mu\psi
  jG^{\mu\nu}\left(A_\nu\times
    I-\frac{1}{2}A_\nu\times\eta\right)\eta^a\right]
e^{\frac{i}{\hbar}j\eta}\nonumber\\  
&=&\epsilon\int djd\eta N^{-1}(j)\eta^{\mu\nu}
\left[\partial_\mu\psi\left(-j\frac{1}{2}A_\nu\times \eta\right)- 
\partial_a\partial_\mu\psi(j\left(A_\nu\times I\right))
\eta^a\right]e^{\frac{i}{\hbar}j\eta}\nonumber\\
&&+O(\epsilon^2), \label{eq:I2}
\end{eqnarray}
where $G^{\mu\nu}$ has been replaced by $\eta^{\mu\nu}$ to keep
only the first order in $\epsilon$. And the first term in
Eq.~(\ref{eq:I2}) does not contribute, i.e., 
\begin{eqnarray}
&&\epsilon\int djd\eta N^{-1}(j)\eta^{\mu\nu}\partial_\mu\psi
\left(-j\left(\frac{1}{2}A_\nu\times
    \eta\right)\right)e^{\frac{i}{\hbar}j\eta}\nonumber\\ 
&=&-\epsilon\int djd\eta \left[N^{-1}(j)\eta^{\mu\nu}\partial_\mu
  \psi\frac{\hbar}{i}\frac{\partial}{\partial
    j^a}\left(\frac{1}{2}A_\nu\times j\right)^a+ 
\left(\frac{1}{2}A_\nu\times
  j\right)^a\frac{\hbar}{i}\frac{\partial}{\partial j^a}
(N^{-1}(j)\eta^{\mu\nu})\right]e^{\frac{i}{\hbar}j\eta}\nonumber\\ 
&=&0+O(\epsilon^2).
\end{eqnarray}
Finally, the only non-vanishing part in $I_2$ is
\begin{eqnarray}
I_2&=&-\epsilon\int djd\eta N^{-1}(j)\partial_a\partial_\mu\psi
\eta^{\mu\nu}( j\left(A_\nu\times
  I\right))\eta^ae^{\frac{i}{\hbar}j\eta}\nonumber\\ 
&=&\epsilon\int djd\eta N^{-1}(j)\partial_a\partial_\mu\psi \eta^{\mu\nu}
\frac{\hbar}{i}\frac{\partial}{\partial j^a} 
( j\left(A_\nu\times I\right))e^{\frac{i}{\hbar}j\eta}\nonumber\\ 
&=&\epsilon\int djd\eta N^{-1}(j)\partial_a\partial_\mu\psi
\eta^{\mu\nu} 
\frac{\hbar}{i}(A_\nu\times I)^ae^{\frac{i}{\hbar}j\eta}\nonumber\\ 
&=&\epsilon N^{-1}(0)\partial_a\partial_\mu\psi
\frac{\hbar}{i}(A^\mu\times I)^a.
\end{eqnarray}
By the same reason as for $I_2$, in $I_3$ only for the non-zero
first order term in $\epsilon$, only terms in the same power of
$\eta$ and $j$ apart from $N^{-1}(j)$ are kept, 
\begin{eqnarray}
I_3&=&-i\frac{\epsilon}{2\hbar}\int djd\eta N^{-1}(j)
\left(j\left(A_\mu\times I\right) j\left(A^\mu\times I\right)
+\frac{1}{4}j\left(A_\mu\times \eta\right) j\left(A^\mu\times
  \eta\right) 
-j\left(A_\mu\times I\right) j\left(A^\mu\times
  \eta\right)\right)\nonumber\\ 
&&\qquad\qquad\qquad\qquad\times\left(\psi-\partial_a\psi\eta^a +
  \frac{1}{2}\partial_a\partial_b\psi\eta^a\eta^b\right)  
e^{\frac{i}{\hbar}j\eta}\nonumber\\
&=&-i\frac{\epsilon}{2\hbar}\int djd\eta N^{-1}(j)
\left[\frac{1}{4}j\left(A_\mu\times \eta\right) 
j\left(A^\mu\times \eta\right)\psi- j\left(A_\mu\times I\right)
j\left(A^\mu\times
  \eta\right)(-\partial_a\psi\eta^a)\right.\nonumber\\ 
&&\left.\qquad\qquad\qquad\qquad\qquad\qquad\qquad+
j\left(A_\mu\times I\right) j\left(A^\mu\times
  I\right)\left(\frac{1}{2}\partial_a\partial_b\psi
  \eta^a\eta^b\right)\right]e^{\frac{i}{\hbar}j\eta}\nonumber\\  
&=&I_{31}+I_{32}+I_{33}.
\end{eqnarray}
We calculate $I_3$ in three parts, $I_{31}$, $I_{32}$ and
$I_{33}$. 
\begin{eqnarray}
I_{31}&=&-\frac{i\epsilon}{2\hbar}\int djd\eta
N^{-1}(j)\frac{1}{4}\psi j\left(A_\mu\times \eta\right)
j\left(A^\mu\times \eta \right)e^{\frac{i}{\hbar}j\eta}\nonumber\\ 
&=&-\frac{i\epsilon}{2\hbar}\int djd\eta N^{-1}(j)\frac{1}{4}\psi
e^{\frac{i}{\hbar}j\eta}\frac{\hbar}{i}\frac{\partial}{\partial
  j^a}\frac{\hbar}{i}\frac{\partial}{\partial
  j^b}\left[(A_\mu\times j)^a (A^\mu\times j)^b\right]\nonumber\\ 
&=&\frac{i\epsilon\hbar}{2}N^{-1}(0)\frac{1}{4}(f_{acb}A^c_\mu)
\eta^{\mu\nu}(f_{bea}A^e_\nu)\psi\nonumber\\ 
&=&\frac{i\epsilon\hbar}{8}N^{-1}(0)\mbox{Tr}(A\cdot A)\psi.
\end{eqnarray}
\begin{eqnarray}
I_{32}&=&-\frac{i\epsilon}{2\hbar}\int djd\eta
N^{-1}(j)[(-j\left(A_\mu\times I\right)) (j\left(A^\mu\times
  \eta\right))(-\partial_a\psi\eta^a)]
e^{\frac{i}{\hbar}j\eta}\nonumber\\  
&=&-\frac{i\epsilon}{2\hbar}\int djd\eta
N^{-1}(j)(-\partial_a\psi)e^{\frac{i}{\hbar}j\eta}\frac{\hbar}{i}
\frac{\partial}{\partial 
  j^a}\frac{\hbar}{i}\frac{\partial}{\partial
  j^b}[(-j\left(A_\mu\times I\right)) (-A^\mu\times
j)^b]\nonumber\\ 
&=&\frac{i\epsilon\hbar}{2}N^{-1}(0)(-\partial_a\psi)
[-(A_\mu\times I)^b\eta^{\mu\nu} (+f_{acb}A^c_\nu)]\nonumber\\ 
&=&\frac{i\epsilon\hbar}{2}N^{-1}(0)(A^\mu\times I)^bf_{acb}
A^c_\mu\partial_a\psi. 
\end{eqnarray}
\begin{eqnarray}
I_{33}&=&-\frac{i\epsilon}{2\hbar}\int djd\eta
N^{-1}(j)j\left(A_\mu\times I\right) j\left(A^\mu\times I\right) 
\left(\frac{1}{2}\partial_a\partial_b\psi\eta^a\eta^b\right)
e^{\frac{i}{\hbar}j\eta}e^{\frac{i}{\hbar}j\eta}\nonumber\\
&=&-\frac{i\epsilon}{2\hbar}\int djd\eta N^{-1}(j)
\left(\frac{1}{2}\partial_a\partial_b\psi\right)
\frac{\hbar}{i}\frac{\partial}{\partial 
  j^a}\frac{\hbar}{i}\frac{\partial}{\partial
  j^b}[(j\left(A_\mu\times I\right)) (j\left(A^\mu\times
  I\right))]\nonumber\\ 
&=&-\frac{i\epsilon\hbar}{2}N^{-1}(0)\left(
  \frac{1}{2}\partial_a\partial_b\psi\right)
\frac{2}{-1}(A_\mu\times I)^a (A^\mu\times I)^b\nonumber\\ 
&=&\frac{i\epsilon\hbar}{2}N^{-1}(0)(A_\mu\times I)^a (A^\mu\times 
I)^b\partial_a\partial_b\psi. 
\end{eqnarray}
The last integral $I_4$ is straightforward.
\begin{equation}
I_4=\frac{i\epsilon\hbar}{2}\int djd\eta
N^{-1}(j)G^{\mu\nu}\partial_\mu\partial_\nu\psi
e^{\frac{i}{\hbar}j\eta}=\frac{i\epsilon\hbar}{2}N^{-1}(0) 
\partial^2\psi.  
\end{equation} 



\end{document}